\documentclass[12pt,reqno]{article}
\usepackage{amsfonts}
\usepackage{amsmath}
\usepackage{amsbsy}

\usepackage{amssymb,latexsym}
 \numberwithin{equation}{section}

\begin{document}
 \allowdisplaybreaks[1]
\title{Braneworld Gravity in Symmetric Space Bulk}
\author{Nejat T. Y$\i$lmaz\\
Department of Electrical and Electronics Engineering,\\
Ya\c{s}ar University,\\
Sel\c{c}uk Ya\c{s}ar Kamp\"{u}s\"{u}\\
\"{U}niversite Caddesi, No:35-37,\\
A\u{g}a\c{c}l\i Yol, 35100,\\
Bornova, \.{I}zmir, Turkey.\\
\texttt{nejat.yilmaz@yasar.edu.tr}} \maketitle
\begin{abstract}
By considering the p-brane motion in $G/K$ symmetric space bulk we
identify the $G$-invariant bulk metric in the solvable lie algebra
gauge of the brane action. After calculating the Levi-Civita
connection of this bulk metric we use it in the Gauss equation to
compute the braneworld curvature in terms of the bulk coordinates.
Finally, by making use of the Gauss equation in the braneworld
Einstein equation we present a geometrical method of implementing
the first fundamental form in the gravitating brane dynamics for
the specially chosen symmetric space bulk case leading to an
Einstein equation solely expressed in terms of the bulk
coordinates of the braneworld.
\end{abstract}
\section{Introduction}
Following \cite{ran1,ran2}, braneworld cosmological scenarios have
gained an extensive attention. Within the last decade there is a
vast literature formed in this direction. As introductory samples
of this literature the reader may consult \cite{bran1} for brane
cosmology in general, \cite{bran2,bran3,bran4,bran5} for inflation
emerging from braneworld scenarios, and \cite{grav1} for
braneworld gravity.

In this work, we study the braneworld gravity when the braneworld
is immersed in a generic symmetric space bulk so that the brane
moves in any sort of symmetric space \cite{hel}. We will focus on
the Einstein equation when the induced braneworld metric is
coupled to the Einsteinian gravity in the presence of other
braneworld matter fields. As our major point of view in this work
is the brane motion in symmetric space bulk we will specify the
generic p-brane action
\cite{pbrane1,pbrane2,pbrane3,pbrane4,pbrane5,pbrane6,pbrane7,pbrane8,howe,polyakov}
so that it will exhibit certain global and local symmetries
emerging from the symmetric space bulk. For this reason we will
refer to the symmetric space sigma model being a coset sigma model
which can be obtained from the principal chiral model by reduction
\cite{sm1,sm2,sm3,Eichen,zak,uhlen,manas}. To implement the above
mentioned symmetries in the p-brane dynamics we will refer to the
equivalent Polyakov action. Basically, we will make use of the
sigma model action for the symmetric space target manifolds in
particular the one constructed in the solvable lie algebra gauge
\cite{julia1,julia2,gauged,nej1,nej2,nej3,sssm1}. The outstanding
characteristics of the brane motion in symmetric space bulk
obeying certain symmetries will appear as the predetermination of
the bulk metric at the braneworld-bulk intersection in terms of
the braneworld coordinates. We will compute the Levi-Civita
connection of the bulk which is compatible with this metric.
Later, by plugging in the ingredients of the derived bulk
curvature in the Gauss equation of the braneworld immersion we
will present the formal method of relating the bulk and the
braneworld geometries. Consequently, we will see that in the last
section, such a relation will help us to implement the induced
metric constraint in the braneworld Einstein equation in a way
which makes the geometry of the immersion explicit and accessible
at the level of field equations.

In Section two, we will give introductory remarks for the free
p-brane motion in a symmetric space bulk. We will also mention
about the equivalence of the Nambu-Goto and the Polyakov actions.
The solvable lie subalgebra gauge construction of the symmetric
space sigma model will be discussed in Section three. Section
four, is reserved for the identification of the bulk metric which
will be read through the sigma model lagrangian and which is
dictated by the symmetries of the gauge set in Section three. In
Section five, we will compute the Levi-Civita connection one-forms
and the curvature two-forms of this bulk metric. In Section six,
we will construct the Gauss equation of the braneworld immersion.
Finally, in the last section we will combine all our machinery of
the previous sections to write down the Einstein equation of the
braneworld in terms of the braneworld coordinates when its induced
metric is coupled to the braneworld gravity and other braneworld
matter fields.
\section{Free p-branes in Symmetric Spaces}\label{section1}
In this section, we will focus on the general properties of the
Nambu-Goto and the equivalent Polyakov actions corresponding to
the free p-brane motion in a symmetric space bulk. Now consider a
Lie group $G$ which is a non-compact real form of any other
semi-simple Lie group and also consider a maximal compact subgroup
of $G$ which we will denote by $K$. If we assume that the Lie
algebra $k$ of $K$ is a maximal compactly imbedded Lie subalgebra
of the Lie algebra of $G$ then it is an element of a Cartan
decomposition of the Lie algebra of $G$. In this case $(G,K)$ is a
Riemannian symmetric pair therefore the left coset space $G/K$ has
a unique analytical structure induced by the quotient topology of
$G$. The manifold $G/K$ is a Riemannian globally symmetric space
for all the $G$-invariant Riemannian structures on $G/K$
\cite{hel}. The coset manifold $G/K$ becomes a homogeneous space
since for all the $G$-invariant Riemannian structures on $G/K$ the
identity component $I_{0}(G/K)$ of the isometry group acts
transitively on $G/K$.

If we consider the low energy limit motion of Dirichlet p-branes
\cite{pbrane1,pbrane2,pbrane3,pbrane4,pbrane5,pbrane6,pbrane7,pbrane8}
in a generic above mentioned $G/K$ symmetric space background the
action which governs the dynamics is the Dirac-Born-Infeld action
\cite{born,dirac}
\begin{equation}\label{e1}
 S_{DBI}=-T_{p}\, \int d^{(p+1)}\sigma\sqrt{-\text{det}(G_{AB}+\mathcal{F}_{AB})},
\end{equation}
where
\begin{equation}\label{e2}
 G_{AB}=g_{ab}\partial_{A}\varphi^{a}\partial_{B}\varphi^{b},
\end{equation}
is the pullback of the semi-Riemannian metric $g_{ab}$ which we
assign on the symmetric space $G/K$ through the immersion map
\cite{darling}
\begin{equation}\label{e3}
f:\:N\longrightarrow G/K,
\end{equation}
of the world volume $N$ of the p-brane in the bulk $G/K$. $T_{p}$
is the p-brane tension. Locally the immersion is characterized by
the coordinates $\varphi^{b}(x^{A})$ for $b=1,...,$dim$(G/K)$ of
the symmetric space bulk $G/K$ which are functions of the local
world volume coordinates $x^{A}$ for $A=1,...,(p+1)$. In
\eqref{e1}
\begin{equation}\label{e4}
\mathcal{F}_{AB}=F_{AB}-B_{AB},
\end{equation}
where $F$ is the field strength of a $U(1)$ gauge field living on
the world volume $N$ and
\begin{equation}\label{e5}
B_{AB}=B_{ab}\partial_{A}\varphi^{a}\partial_{B}\varphi^{b},
\end{equation}
is the pullback of a two-form field living on $G/K$ onto the world
volume $N$. Now if we set
\begin{equation}\label{e6}
\mathcal{F}_{AB}=0,
\end{equation}
we obtain the Nambu-Goto action
\begin{equation}\label{e7}
 S_{NG}=-T_{p}\, \int d^{(p+1)}\sigma\sqrt{-\text{det}(G_{AB})},
\end{equation}
for the free p-brane which moves in the symmetric space $G/K$
bulk. Now if we consider the world volume $N$ as an isometrically
immersed submanifold of the semi-Riemannian manifold $G/K$ which
is endowed with $g_{ab}$ then \eqref{e2} becomes the first
fundamental form of the immersion
\cite{aminov,docarmo,sternberg,oneil}. Also \eqref{e7} becomes a
multiple of the semi-Riemannian volume of the world volume $N$. We
should remark that seing the world volume as an isometric
immersion is equivalent to the fact that $G_{AB}$ in \eqref{e2} is
the induced metric on $N$. However in this framework it is not an
independent field. Now let us consider the Polyakov action
\begin{equation}\label{e8}
 S_{P}=-T\, \int d^{(p+1)}\sigma\sqrt{-G}\:G^{AB}
 g_{ab}\partial_{A}\varphi^{a}\partial_{B}\varphi^{b},
\end{equation}
whose independent fields are $\{\varphi^{a},G^{AB}\}$ which live
on $N$. \eqref{e8} can also be written as
\begin{equation}\label{e9}
 S_{P}=-T\, \int
 g_{ab}\:d\varphi^{a}\wedge\ast d\varphi^{b}.
\end{equation}
If we vary \eqref{e9} and equate it to zero then we get
\begin{equation}\label{e10}
f(\delta\varphi^{a})-\delta
e^{A}\wedge[T\,g_{ab}(d\varphi^{b}\wedge i_{A}\ast
d\varphi^{a}+i_{A}d\varphi^{a}\wedge\ast d\varphi^{b})]=0.
\end{equation}
Here $\{e^{A}\}$ is a viel-bein on $N$. Also $i_{A}$ is the
interior derivative with respect to $e^{A}$. \eqref{e10} can
further be written as
\begin{equation}\label{e11}
f(\delta\varphi^{a})-\delta e^{A}\wedge[-2\,
T\,g_{ab}\partial_{A}\varphi^{a}\partial_{B} \varphi^{b}+T\,
g_{ab}\partial^{C}\varphi^{a}\partial_{C}\varphi^{b}G_{AB}]\ast
e^{B}=0,
\end{equation}
where the expression inside the brackets is the energy-momentum
tensor corresponding to the action \eqref{e9}. Thus we have
\begin{equation}\label{e12}
-2g_{ab}\partial_{A}\varphi^{a}\partial_{B}
\varphi^{b}+g_{ab}\partial^{C}\varphi^{a}\partial_{C}\varphi^{b}G_{AB}=0.
\end{equation}
We observe that assuming an isometric immersion solution namely
\eqref{e2} in \eqref{e12} yields
\begin{equation}\label{e13}
-2+p+1=0.
\end{equation}
We see that only for $p=1$ (for strings) \eqref{e2} is a solution
of \eqref{e12}. In this case if one inserts \eqref{e2} in the
Polyakov action \eqref{e8} then one obtains the Nambu-Goto action
\eqref{e7} with $T=T_{p}/2$. Thus both of the actions are
equivalent when one considers the field equations of
$\{\varphi^{a}\}$. However for $p>1$ the equivalence of \eqref{e7}
and \eqref{e8} is obviously not possible. Even if one considers
\eqref{e8} as a constraint system with a constraint equation
$G_{AB}\:\propto\:g_{ab}\partial_{A}\varphi^{a}\partial_{B}\varphi^{b}$
to establish the equivalence such a constraint equation is
inconsistent with the field equations \eqref{e12}. Of course the
privileged state of the string case is due to the Weyl-invariance
which occurs only when $p=1$. On the other hand when $p>1$ one may
introduce a cosmological constant term to \eqref{e8} to establish
a similar equivalence \cite{howe,polyakov}. For $p>1$ one should
consider the action
\begin{equation}\label{e14}
 S_{P}=-T\, \int d^{(p+1)}\sigma\sqrt{-G}\:G^{AB}
 g_{ab}\partial_{A}\varphi^{a}\partial_{B}\varphi^{b}-T\int d^{(p+1)}\sigma
 \sqrt{-G}\:\Lambda,
\end{equation}
which can also be written as
\begin{equation}\label{e15}
 S_{P}=-T\, \int
 g_{ab}\:d\varphi^{a}\wedge\ast d\varphi^{b}-T\int\ast\Lambda.
\end{equation}
Following the same track above the corresponding Einstein equation
yields
\begin{equation}\label{e16}
-2g_{ab}\partial_{A}\varphi^{a}\partial_{B}
\varphi^{b}+g_{ab}\partial^{C}\varphi^{a}\partial_{C}\varphi^{b}G_{AB}+\Lambda
G_{AB}=0.
\end{equation}
We see that if we chose $\Lambda=1-p$ then \eqref{e2} satisfies
this equation and again the Polyakov \eqref{e14} and the
Nambu-Goto \eqref{e7} actions become equivalent for the fields
$\{\varphi^{a}\}$ following the insertion of \eqref{e2} in
\eqref{e14}. Also in this case $T=T_{p}/2$. When $p\neq 1$ the
solution of \eqref{e16} is
\begin{equation}\label{e17}
 G_{AB}=-\frac{p-1}{\Lambda}g_{ab}\partial_{A}\varphi^{a}\partial_{B}\varphi^{b}.
\end{equation}
Thus if one wishes to leave $\Lambda$ arbitrary and still
establish the equivalence of \eqref{e14} and \eqref{e7} via
insertion of \eqref{e17} in \eqref{e14} then one must give up the
isometric immersion condition \eqref{e2} by scaling it. In this
case by assuming positive $\Lambda$ we find
$T_{p}=-2T(\frac{\Lambda}{p-1})^{\frac{1-p}{2}}$.
\section{Symmetries and the Gauge}\label{section1}
We have reviewed the conditions on the equivalence of the free
p-brane and the sigma model actions in the last section. Now
bearing in mind that our perspective is the brane motion in
symmetric space bulk we will refine the sigma model action so that
it will admit global and local symmetries. Equivalently this
corresponds to restricting ourselves to brane motion with a
collection of local and global conserved currents and charges
which govern the dynamics of the brane in the bulk. We will
shortly review the construction of the $G/K$ symmetric space sigma
model and then establish the solvable Lie algebra gauge to
finalize the form of the Polyakov action which we will refer to
for the rest of our analysis. Here $G$ is a real form of a
non-compact semisimple Lie group and $K$ is a maximal compact
subgroup of $G$. To construct a Lagrangian which has global and
local symmetries one starts from the $G$-valued field $\nu(x)\:
:N\longrightarrow G$ and considers the pull-back $\mathcal{G}=\nu
d\nu^{-1}$ of the Cartan-Maurer form of the Lie group $G$ in a
matrix representation which can be decomposed as
\begin{equation}\label{e18}
\mathcal{G}=\mathcal{G}_{A}dx^{A}=Q_{A}^{a}(x)K_{a}dx^{A}+P_{A}^{b}(x)T_{b}dx^{A}=Q+P.
\end{equation}
Here  $\{K_{a}\}$ are the generators of the Lie algebra of $K$ and
$\{T_{b}\}$ form a basis for its vector space direct sum
complement $p$ in the Lie algebra of $G$. These definitions are
based on the Cartan decomposition of the Lie algebra $g$ of $G$
\begin{equation}\label{e19}
g=k\oplus p,
\end{equation}
where $k$ is the Lie algebra of $K$ and $p$ is its complement in
$g$. The Lie algebra $k$ is a maximal compactly imbedded Lie
subalgebra of $g$. Also $k$ and $p$ are orthogonal with respect to
the trace of a representation chosen. This is a consequence of the
homogeneous space nature of $G/K$ which enables the existence of
an inner product on $g$ which is adjoint-invariant and which is
positive definite on $p$. This inner product can also be projected
onto $G/K$. If one chooses the adjoint representation then it
becomes the Cartan-Killing form. Thus $tr(kp)=0$. As \eqref{e19}
is a Cartan decomposition it admits
\begin{equation}\label{e19.5}
[k,k]\subset k\quad,\quad [k,p]\subset p\quad,\quad [p,p]\subset
k,
\end{equation}
where the last commutation relation quarantines that \eqref{e19}
is a Cartan decomposition and $G/K$ is a symmetric space
\cite{hel} which is a subclass of homogeneous spaces. By adding
this commutation relation to the first two one reduces the
homogeneous space sigma model to the symmetric space sigma model
which admits a Lax pair thus becomes an integrable system possing
an infinite number of local conserved charges. This specialization
will also enable us to introduce an involutive automorphism called
the generalized transpose which leads to the internal metric
formulation of the associated sigma model. Now if we define
$P_{A}\equiv P_{A}^{a}T_{a}$ then the sigma model action which has
$G$-global invariance from the right ($\nu\longrightarrow \nu
g^{\prime}\quad \forall g^{\prime}\in G $) and $K$-local
invariance from the left ($\nu\longrightarrow
k^{\prime}(x)\nu\quad \forall k^{\prime}(x)\in K $ and $x\in N$)
can be constructed as \cite{sch1}
\begin{equation}\label{e20}
S_{Sigma}=T\, \int
d^{(p+1)}\sigma\sqrt{-G}\:tr(P_{A}P^{A})=T\,\int tr(\ast P\wedge
P).
\end{equation}
In \eqref{e18} $Q$ appears as a gauge connection corresponding to
the local $K$-symmetry thus one may define the covariant
derivative
\begin{equation}\label{e20.5}
D\nu=d\nu+Q\nu,
\end{equation}
which in component form can be written as
$D_{A}\nu=\partial_{A}\nu+Q_{A}\nu$. Thus under the local
$K$-symmetry we can identify the covariant part of \eqref{e18} as
\begin{equation}\label{e20.6}
P=-D\nu\nu^{-1}.
\end{equation}
On the other hand $\mathcal{G}=\nu d\nu^{-1}$ is the corresponding
Noether current of the global $G$-symmetry and both $Q$ and $P$
are invariant under the global $G$-right-action. By using
\eqref{e20.6} the Lagrangian \eqref{e20} can also be written as
\begin{equation}\label{e20.7}
S_{Sigma}=T\, \int tr(\ast D\nu\nu^{-1}\wedge D\nu\nu^{-1}).
\end{equation}
The Cartan decomposition in \eqref{e19} is the eigenspace
decomposition of an involutive automorphism of $g$ which is called
the Cartan involution. On the components based on the
decomposition \eqref{e19} it acts as
\begin{equation}\label{e20.8}
S\: :\: (Q,P)\longrightarrow (Q,-P).
\end{equation}
By using this involution we can define a generalized transpose
operator $\#$ on $g$ as
\begin{equation}\label{e20.85}
\#(g^{\prime})=-S(g^{\prime})\quad \forall g^{\prime}\in g.
\end{equation}
Since it is induced by the Cartan involution $S$ one is able to
find a higher dimensional matrix representation of $g$ in which
$\#$ coincides with the ordinary matrix transpose operator. For
this reason one can extend \eqref{e20.85} to the group $G$ and one
can generally define an induced map $\#$ over $G$ as
$(exp(g))^{\#}=exp(g^{\#})$. In this respect both \eqref{e20.85}
and $(exp(g))^{\#}=exp(g^{\#})$ are representation-free and they
admit all the features of the matrix transpose operator justifying
their name. By using this operator we can write $P$ as
\begin{equation}\label{e20.86}
P=\frac{1}{2}(\nu d\nu^{-1}+(\nu d\nu^{-1})^{\#}).
\end{equation}
Now if we define
\begin{equation}\label{e20.87}
\mathcal{M}=\nu^{\#}\nu,
\end{equation}
and further
\begin{equation}\label{e20.88}
\mathcal{A}=\mathcal{M}^{-1}d\mathcal{M},
\end{equation}
then by using \eqref{e20.86} we can show that
\begin{equation}\label{e20.89}
\mathcal{A}=-2\nu^{-1}P\nu.
\end{equation}
$\mathcal{M}$ provides a parametrization of the coset space $G/K$.
Since its definition resembles the metric construction from the
viel-bein it is usually called the internal metric. Now by using
\eqref{e20.89} in \eqref{e20} and the fact that trace operator
permits cyclic permutations we can write
\begin{equation}\label{e20.891}
S_{Sigma}=T\,\int tr(\ast P\wedge
P)=\frac{T}{4}tr(\ast\mathcal{A}\wedge\mathcal{A})=-\frac{T}{4}tr(\ast
d\mathcal{M}\wedge d\mathcal{M}),
\end{equation}
where in the last identity we have used
$d\mathcal{M}^{-1}\mathcal{M}=-\mathcal{M}^{-1}d\mathcal{M}$. Due
to the local gauge symmetry the solution space $\{\nu(x)\}$ of
\eqref{e20.891} has non-physical gauge degrees of freedom which
can be eliminated by choosing a gauge and omitting the gauge
symmetry. An appropriate gauge which is entirely based on the
physical degrees of freedom is the solvable Lie algebra gauge
\cite{fre}. The solvable Lie algebra gauge fixed field is
\begin{equation}\label{e21}
\nu (x)=e^{\frac{1}{2}\phi ^{i}(x)H_{i}}e^{\chi
^{\beta}(x)E_{\beta}},
\end{equation}
which is based on the axion-dilaton parametrization of the
symmetric space $G/K$ \cite{nej1,nej2,nej3,sssm1,symmspace}. The
map
\begin{equation}\label{e23}
Exp:p\longrightarrow G/K,
\end{equation}
where $p$ is the complementing piece in the Cartan decomposition
\eqref{e19} is a diffeomorphism onto $G/K$ from $p$ which can be
considered as a submanifold of $g$ that can be identified with the
tangent space of $G$ at the identity element. Now if we turn back
to \eqref{e21} the right hand side is composed of two factors of
the exponential map of the Lie group $G$. $\{H_{i}, E_{\beta}\}$
are the generators of a solvable Lie subalgebra $s$ of $g$ which
takes part in the Iwasawa decomposition
\begin{equation}\label{e24}
g=k\oplus s,
\end{equation}
of the Lie algebra $g$ \footnote{$\{H_{i}\}$ are the non-compact
Cartan generators and $\{E_{\beta}\}$ are a subset of the positive
root generators which correspond to the non-compact positive roots
of the Lie algebra of $G$ with respect to the Cartan involution.
They form a nilpotent Lie subalgebra of $g$.} \cite{hel}. We
should state here that the decompositions \eqref{e19} and
\eqref{e24} are not the same. Furthermore as a vector space
\begin{equation}\label{e25}
s\cong p,
\end{equation}
thus
\begin{equation}\label{e26}
Exp:s\longrightarrow G/K,
\end{equation}
is another diffeomorphism from $s$ which can be equipped with the
differentiable structure of $p$ onto $G/K$. Therefore \eqref{e26}
which the gauge \eqref{e21} refers to is a global parametrization
of $G/K$. We can use this new gauge \eqref{e21} which contains
only the true degrees of freedom of the symmetric space sigma
model in  \eqref{e20.87}. Thus when one fixes the gauge
\eqref{e21} the gauge fixed solutions of \eqref{e20} can be
obtained from the action
\begin{equation}\label{e27}
 S_{Sigma}=-\frac{T}{4}\, \int tr(\ast d{\mathcal{M}}^{-1}\wedge
 d{\mathcal{M}}).
\end{equation}
Since we have chosen a gauge and eliminated the gauge degrees of
freedom of the sigma model \eqref{e27} breaks the local
$K$-symmetry but it is still $G$-global invariant. However the
standard global right action of $G$ on $G$ namely
$\nu\longrightarrow \nu g^{\prime}\quad \forall g^{\prime}\in G $
is no more consistent with the gauge \eqref{e21}. In the gauge
fixed case the global right action of $G$ which preserves the
gauge (whose image always maps $e^{s}$ onto $e^{s}$) can be
defined as
\begin{equation}\label{e28}
\nu\longrightarrow k^{\prime}(g^{\prime},\nu)\nu g^{\prime},\quad
\forall g^{\prime}\in G,\quad \text{and where}\quad k^{\prime}\in
K.
\end{equation}
Under this global action which owes its consistency of closure on
the image of $e^{s}$ to the Iwasawa decomposition \eqref{e24} the
action \eqref{e27} remains invariant. We remark that in
\eqref{e28} one has to introduce a field dependent
$k^{\prime}(g^{\prime},\nu)$ factor from the left to restore the
gauge by pulling the acted field back into the image of $e^{s}$ as
it is thrown out of the gauge by the right factor in
\eqref{e28}\footnote{This is needed to be able to realize the
action of $G$ which is originally defined for the map $\nu$ on the
dilatons and the axions}. Of course via \eqref{e26} this action
also induces an action on the bulk $G/K$ since each group element
$[e^{s^{\prime}}]$ for $s^{\prime}\in s$ is a representative of
the left coset space $G/K$. We may say that the global action of
$G$ on $G/K$ is non-linearly realized through \eqref{e28}. The
bulk metric on $G/K$ is implicitly implemented in \eqref{e27} and
it is dictated by the gauge \eqref{e21} and the global and the
local symmetries. Before passing to the identification of the
explicit form of this bulk metric which may be considered to live
just necessarily at the braneworld occupation of the bulk we will
mention a couple of points about the global axion-dilaton
parametrization of the bulk emerging from \eqref{e21}. Now we have
\begin{equation}\label{e29}
\{\varphi^{a}\}=
  \{\phi^{1},
  \phi^{2},
  \cdots,
  \phi^{r},
  \chi^{1},
  \chi^{2},
  \cdots,
  \chi^{n}\},
\end{equation}
where $r$ is the number of the non-compact Cartan generators in
the Iwasawa decomposition \eqref{e24}, and $n$ is the dimension of
the nilpotent component of the solvable Lie subalgebra $s$ of $g$.
Since in \eqref{e21} we have established the gauge solely with the
physical degrees of freedom their sum is
\begin{equation}\label{e30}
 r+n=\text{dim}(G/K).
\end{equation}
Although \eqref{e26} is a global parametrization of $G/K$ our
point of view will be local and we will assume that the global
parametrization \eqref{e26} coincides with a local coordinate
chart of $G/K$. For this reason we assume that the map
\begin{equation}\label{e31}
C: (\varphi^{1},\varphi^{2},\cdots ,\varphi^{dim(s)})\in
\Bbb{R}^{dim(s)}\longrightarrow (\phi ^{i}H_{i}+\chi
^{\beta}E_{\beta})\in s,
\end{equation}
is a homeomorphism on an open set $U$ of $\Bbb{R}^{dim(s)}$. Thus
the map
\begin{equation}\label{e31}
C^{\prime}=Exp\circ C\: :\: \Bbb{R}^{dim(s)}\longrightarrow G/K,
\end{equation}
is also a homeomorphism and $(C^{\prime}(U),C^{\prime\:-1})$
becomes a coordinate chart for $G/K$ and $\{\varphi^{a}\}$ become
the coordinates of $G/K$.\footnote{We should state that
$dim(p)=dim(s)=dim(G/K)$.} On this coordinate chart by also
assuming a local chart for $N$ we may take the gauge fixed
Polyakov action that is equivalent to \eqref{e7} as
\begin{equation}\label{e32}
 S_{p}=-\frac{T}{4}\, \int tr(\ast d{\mathcal{M}}^{-1}\wedge
 d{\mathcal{M}})-T\int\ast\Lambda,
\end{equation}
which is invariant under the global action \eqref{e28} of $G$ on
the bulk; the sigma model target manifold. At first glance this
refined action with the global symmetry \eqref{e28} has invariant
Noether currents which are intimately related to the Cartan forms
$\mathcal{G}=\nu^{-1}d\nu$ and
$\mathcal{G}^{\prime}=d\nu\nu^{-1}$. They can be expressed in
terms of the brane coordinates in the bulk. Furthermore as a
symmetric space sigma model the sigma model term in \eqref{e32}
carries the rich class of characteristic symmetries of the
integrable systems.
\section{$G$-invariant Metric on the Bulk}
As we have mentioned above choosing the gauge \eqref{e21} has
eliminated the non-physical degrees of freedom in the field
$\nu(x)$. Besides together with the global symmetry \eqref{e28} it
also dictates the bulk metric in \eqref{e32}. In this section we
will identify this bulk metric. Its form is a consequence of the
global and the local symmetries of \eqref{e20}, the solvable Lie
algebra gauge \eqref{e21}, and the gauge fixed form of the sigma
model term in \eqref{e32} which possesses the global symmetry
\eqref{e28}. Another gauge may result in a different bulk metric
however the physical solution space will be the same. One may
argue the physical meaning of this gauge-dependent bulk metric.
First of all we should see that since in \eqref{e32} the bulk
coordinates depend on the world volume ones it determines the
metric at the regions of the bulk traced by the brane. With this
point of view we may call this a moving metric that is to say it
is the metric required at the braneworld to establish the
necessary symmetries of the motion incorporation with the gauge
chosen. We may consider it as a constraint on the metric structure
of the bulk even if one introduces bulk gravity. This constraint
emerges from the symmetries of the brane motion we require. Each
gauge fixing may end up with a different constraint system whose
physical solution space of the brane coordinates will be the same.
Now let us focus on \eqref{e32}. In \cite{sssm1} the kinetic term
of \eqref{e32} is explicitly derived in terms of the bulk
coordinates $\{\varphi^{a}\}$. It reads
\begin{subequations}\label{e33}
\begin{align}
-\frac{T}{4}\, \int tr(\ast d{\mathcal{M}}^{-1}\wedge
 d{\mathcal{M}})&=-T\int\bigg(-\frac{1}{8}\,\mathcal{A}_{ij}\ast d{\phi}^{i}\wedge
 d\phi^{j}\notag\\
 &\quad-\frac{1}{4}\,\mathcal{B}_{i\alpha}\ast d{\phi}^{i}\wedge
e^{\frac{1}{2}\alpha_{j}\phi^{j}} \mathbf{\Omega}^{\alpha}_{\gamma}d\chi^{\gamma}\notag\\
&\quad-\frac{1}{2}\,\mathcal{C}_{\alpha\beta}e^{\frac{1}{2}\alpha_{j}\phi^{j}}\ast
\mathbf{\Omega}^{\alpha}_{\gamma}d\chi^{\gamma}\wedge
e^{\frac{1}{2}\beta_{i}\phi^{i}}
\mathbf{\Omega}^{\beta}_{\tau}d\chi^{\tau}\bigg) ,\tag{\ref{e33}}
\end{align}
\end{subequations}
where the coefficients
$\mathcal{A}_{ij},\mathcal{B}_{i\alpha},\mathcal{C}_{\alpha\beta}$
are the normalization constants of the generators $\{H_{i},
E_{\beta}\}$ in the matrix representation chosen for the Lie
algebra of $G$ namely
\begin{subequations}
\begin{gather}\label{e34}
\mathcal{A}_{ij}=tr(H_{i}H_{j}^{\#})+tr(H_{i}H_{j}),\notag\\
\notag\\
\mathcal{B}_{i\alpha}=tr(H_{i}E_{\alpha}^{\#})+tr(E_{\alpha}H_{i}^{\#})+
tr(H_{i}E_{\alpha})+tr(E_{\alpha}H_{i}),\notag\\
\notag\\
\mathcal{C}_{\alpha\beta}=tr(E_{\alpha}E_{\beta}^{\#})+tr(E_{\alpha}E_{\beta})\tag{\ref{e34}}.
\end{gather}
\end{subequations}
 $\alpha_{j}$ and $\beta_{i}$ are the root vector components of the
Cartan generators $\{H_{i}\}$. Their definitions may be referred
in \cite{sssm1}. On the other hand \eqref{e33} introduces the
functions
\begin{equation}\label{e35}
\mathbf{\Omega}= \mathbf{\Omega}(\chi^{\beta}),
\end{equation}
which are rigorously derived in the references
\cite{nej1,nej2,nej3,symmspace}. Briefly $\mathbf{\Omega}$ is a
dim$n\times$dim$n$ matrix
\begin{equation}\label{e36}
 \mathbf{\Omega}=(e^{\omega}-I)\,\omega^{-1},
\end{equation}
where the matrix $\omega$ is
\begin{equation}\label{e37}
 \omega _{\beta }^{\gamma }=\chi ^{\alpha }\,K_{\alpha \beta
}^{\gamma}.
\end{equation}
The structure constants $K_{\alpha \beta }^{\gamma }$ are defined
as
\begin{equation}\label{e38}
[E_{\alpha },E_{\beta }]=K_{\alpha \beta }^{\gamma }\,E_{\gamma }.
\end{equation}
In \cite{sssm1} it is discussed in detail that if one writes
\eqref{e36} as a series expansion then the series will terminate
after a finite number of terms which eases the explicit
calculation of \eqref{e36}. In \eqref{e33} $i,j$ run from $1$ to
$r$ and $\alpha,\beta,\gamma,\tau$ run from $1$ to $n$
respectively. The field equations of the dilatons $\{\phi^{i}\}$
and the axions $\{\chi^{\beta}\}$ of the action \eqref{e33} are
already derived in \cite{sssm1}. They correspond to the field
equations of $\{\varphi^{b}(x^{A})\}$ which are the compositions
of the bulk coordinates with the world volume ones of the freely
moving p-brane whose equivalent action is given in \eqref{e32}.
Since the cosmological term in \eqref{e32} does not depend on the
fields $\{\varphi^{b}(x^{A})\}$ the field equations derived in
\cite{sssm1} also correspond to the field equations of
\eqref{e32}. In this section, our main perspective is to identify
the $G$-invariant metric on the symmetric space bulk $G/K$ which
is a consequence of the gauge chosen in the last section,
therefore for the field equations of the free brane motion we
refer the reader to \cite{sssm1}. Now following our discussion at
the end of the previous section on the assumed coordinate chart
$(C^{\prime}(U),C^{\prime\:-1})$ if we compare the right hand side
of \eqref{e33} with the first term on the right hand side of
\eqref{e15} we may read the components of the bulk metric
\begin{equation}\label{e39}
g = g_{ab}\:d\varphi^{a}\otimes d\varphi^{b},
\end{equation}
on $G/K$. This metric characterizes the gauge fixed form
\eqref{e27} of the G-global and the K-local symmetric sigma model
in \eqref{e20} which takes part in \eqref{e32} which is the
Polyakov equivalent of the free brane action in symmetric space
bulk. The existence of such a metric is due to the invariant
construction of the sigma model \cite{gauged} and the local
equivalence of the minimal, lifted, viel-bein, and the internal
metric constructions of the sigma model on the coordinate chart
$(C^{\prime}(U),C^{\prime\:-1})$
\cite{julia1,julia2,gauged,west,tanii}. The components of $g$ can
now be given as
\begin{subequations}\label{e40}
\begin{align}
g_{ij}=&-\frac{1}{8}\mathcal{A}_{ij},\notag\\
\notag\\
g_{i,r+\beta}=&-\frac{1}{8}\,\mathcal{B}_{i\alpha}
e^{\frac{1}{2}\alpha_{j}\phi^{j}} \mathbf{\Omega}^{\alpha}_{\beta},\notag\\
\notag\\
g_{r+\beta,i}=&-\frac{1}{8}\,\mathcal{B}_{i\alpha}
e^{\frac{1}{2}\alpha_{j}\phi^{j}} \mathbf{\Omega}^{\alpha}_{\beta},\notag\\
\notag\\
g_{r+\gamma,r+\tau}=&-\frac{1}{2}\,\mathcal{C}_{\alpha\beta}e^{\frac{1}{2}\alpha_{j}\phi^{j}}
e^{\frac{1}{2}\beta_{i}\phi^{i}}\mathbf{\Omega}^{\alpha}_{\gamma}
\mathbf{\Omega}^{\beta}_{\tau}.\tag{\ref{e40}}
\end{align}
\end{subequations}
In matrix form the metric can be written as\footnote{$C$
corresponds to Cartan directions, $M$ corresponds to mixed
directions, and $N$ corresponds to the nilpotent directions on
$G/K$.}
\begin{equation}\label{e41}
g=\left(\begin{array}{cc}
  C_{(r\times r)} & M_{(r\times n)}  \\
  M_{(n\times r)}^{T} & N_{(n\times n)} \\
\end{array}\right),
\end{equation}
where
\begin{equation}\label{e42}
C_{ij}=g_{ij}\quad ,\quad M_{i\beta}=g_{i,r+\beta}\quad ,\quad
M_{\beta i}=g_{r+\beta,i}\quad ,\quad
N_{\gamma\tau}=g_{r+\gamma,r+\tau}.
\end{equation}
We can explicitly write down the metric $g$ on $G/K$ in the
solvable Lie algebra parametrization introduced in the gauge
\eqref{e21} as
\begin{subequations}\label{e43}
\begin{align}
g=&-\frac{1}{8}\mathcal{A}_{ij}d\phi^{i}\otimes
d\phi^{j}-\frac{1}{8}\,\mathcal{B}_{i\alpha}
e^{\frac{1}{2}\alpha_{j}\phi^{j}}
\mathbf{\Omega}^{\alpha}_{\beta}d\phi^{i}\otimes d\chi^{\beta}
-\frac{1}{8}\,\mathcal{B}_{i\alpha}
e^{\frac{1}{2}\alpha_{j}\phi^{j}}
\mathbf{\Omega}^{\alpha}_{\beta}d\chi^{\beta}\otimes
d\phi^{i}\notag\\
&-\frac{1}{2}\,\mathcal{C}_{\alpha\beta}e^{\frac{1}{2}\alpha_{j}\phi^{j}}
e^{\frac{1}{2}\beta_{i}\phi^{i}}\mathbf{\Omega}^{\alpha}_{\gamma}
\mathbf{\Omega}^{\beta}_{\tau}d\chi^{\gamma}\otimes
d\chi^{\tau}.\tag{\ref{e43}}
\end{align}
\end{subequations}
By using the properties of the generalized transpose operator $\#$
which coincides with the ordinary matrix transpose in specially
chosen representations as discussed in \cite{sssm1} one can easily
verify that
\begin{equation}\label{e44}
\mathcal{A}_{ij}=\mathcal{A}_{ji}\quad ,\quad
\mathcal{C}_{\alpha\beta}=\mathcal{C}_{\beta\alpha}.
\end{equation}
Also by rearranging indices and with the help of \eqref{e44} one
can prove that
\begin{equation}\label{e45}
g_{ij}=g_{ji}\quad ,\quad g_{i,r+\beta}=g_{r+\beta,i}\quad ,\quad
g_{r+\gamma,r+\tau}=g_{r+\tau,r+\gamma},
\end{equation}
which justifies the symmetry of $g$
\begin{equation}\label{e46}
g_{ab}=g_{ba}.
\end{equation}
The bi-linearity follows from the construction of $g$ in
\eqref{e43}. On the other hand as we have stated before if we
assume that the axion-dilaton parametrization locally satisfies to
be a coordinate chart $(C^{\prime}(U),C^{\prime\:-1})$ then
\begin{equation}\label{e47}
\{d\varphi^{a}\}=\{d\phi^{i},d\chi^{\beta}\},
\end{equation}
becomes a moving co-frame on the bulk. The dual moving frame being
\begin{equation}\label{e48}
\{\partial_{\varphi^{a}}\},
\end{equation}
satisfies
\begin{equation}\label{e49}
d\varphi^{a}(\partial_{\varphi^{b}})=\delta^{a}_{\; b},
\end{equation}
and we have
\begin{equation}\label{e50}
g_{ab}=g(\partial_{\varphi^{a}},\partial_{\varphi^{b}}).
\end{equation}
Since the local construction of the sigma model via the coordinate
chart $(C^{\prime}(U),C^{\prime\:-1})$ guarantees the existence of
the metric \eqref{e43} the non-degeneracy of it namely
\begin{equation}\label{e51}
\text{If}\quad g(X,Y)=0,\quad \forall X\in
E^{1}(C^{\prime}(U))\;\Longrightarrow\; Y=0,
\end{equation}
assures that
\begin{equation}\label{e52}
det\left(\begin{array}{cc}
  C_{(r\times r)} & M_{(r\times n)}  \\
  M_{(n\times r)}^{T} & N_{(n\times n)} \\
\end{array}\right)\neq 0,
\end{equation}
on $C^{\prime}(U)$ \cite{cornwell,kunze}.
Also as we have mentioned
before the calculation of \eqref{e36} drops to a finite number of
terms and $\mathbf{\Omega}(\chi^{\beta})$ becomes a polynomial
function. Consequently the metric components $g_{ab}$ become
smooth functions for the general parametrization in particular for
the local coordinate chart we assume.

The metric \eqref{e43} locally defines a pseudo-Riemannian
structure on $C^{\prime}(U)$. One may furthermore inspect the
conditions (namely the particular brane motion types) which result
in the positivity of the metric so that it defines a local
Riemannian structure.

We will also briefly mention about the distinct symmetry property
of the metric $g$ in \eqref{e43}. The right action \eqref{e28} of
$G$ on $G/K$ generates induced vector fields on $G/K$ which are in
one to one correspondence with the tangent space $T_{e}G$ at the
identity element or the right-invariant vector fields on $G$.
These induced vector fields are defined as follows; the one
parameter subgroups of $G$ which are the integral curves of the
right-invariant vector fields passing from the identity element of
$G$ generate one parameter group of local diffeomorphisms on $G/K$
via the action of $G$ on $G/K$, the vector fields whose integral
curves coincide with the induced curves of these one parameter
group of local diffeomorphisms are the induced vector fields on
$G/K$. In other words induced vector fields are the ones whose
flows coincide with these action-generated one parameter group of
local diffeomorphisms \cite{isham}. Under Lie brackets the induced
vector fields become a Lie algebra and they form a subalgebra of
the Lie algebra of $G$. Now in order that the kinetic term on the
right hand side of \eqref{e32} is invariant under the global right
action of $G$ on $G/K$ the induced vector fields on $G/K$ must be
the Killing vectors of the metric \eqref{e43} defined on
$C^{\prime}(U)$ \cite{gauged}. That is to say if $\{K_{I}\}$ is a
basis for the induced vector fields then
\begin{equation}\label{e53}
\mathcal{L}_{K_{I}}g=0,
\end{equation}
where $\mathcal{L}$ is the Lie derivative on $G/K$.\footnote{We
can at most speak about the Lie derivative of the metric with
respect to the restrictions of $\{K_{I}\}$ on $C^{\prime}(U)$
where the metric is identified.} We may equivalently state this as
follows; if $\Phi: G\longrightarrow$ Diff$(G/K)$ is the right
action of $G$ on $G/K$ then $\Phi(G)\subset I(g_{ab})\subset$
Diff$(G/K)$ where $I(g_{ab})$ is the isometry group of the bulk
metric $g_{ab}$. Thus for the global invariance of the kinetic
part of the action \eqref{e32} we must have
$\Phi(g^{\prime})^{\ast}g=g\quad\forall g^{\prime}\in G$.
Following this observation we can conclude that the bulk metric
\eqref{e43} on $G/K$ is dictated by the symmetry of the theory
that is to say it is implicitly determined by the global right
action of $G$ on $G/K$.
\section{The Levi-Civita Connection on the Bulk}\label{section2}
In this section, we will calculate the Levi-Civita connection on
the bulk which is compatible with \eqref{e43} and which is the
unique torsion-free connection on the tangent bundle $TM(G/K)$. To
start with, by inspecting the metric \eqref{e43} we will choose a
moving co-frame on $TM(G/K)$. From \eqref{e43} we easily see that
if we choose a moving co-frame on $G/K$ as
\begin{equation}\label{e54}
e^{i}=d\phi^{i}\quad ,\quad
e^{\alpha}=e^{\frac{1}{2}\alpha_{j}\phi^{j}}
\mathbf{\Omega}^{\alpha}_{\beta}d\chi^{\beta},
\end{equation}
where $i,j=1,...,r$ and $\alpha,\beta=1,....,n$ then in this frame
the bulk metric can be expressed as
\begin{equation}\label{e55}
g=-\frac{1}{8}\mathcal{A}_{ij}e^{i}\otimes
e^{j}-\frac{1}{8}\,\mathcal{B}_{i\alpha}e^{i}\otimes
e^{\alpha}-\frac{1}{8}\,\mathcal{B}_{i\alpha}e^{\alpha}\otimes
e^{i}-\frac{1}{2}\,\mathcal{C}_{\alpha\beta}e^{\alpha}\otimes
e^{\beta}.
\end{equation}
We should observe that in the moving co-frame \eqref{e54} the
metric components become constant
\begin{equation}\label{e56}
dg_{ab}=0.
\end{equation}
This will bring a major simplification in the calculation of the
corresponding pseudo-Riemannian connection. We realize that the
elements of the moving co-frame in \eqref{e54} are proportional to
the coefficients of the Cartan form ${\mathcal{G}}=d\nu \nu ^{-1}$
which is explicitly calculated in terms of the axions and the
dilatons in \cite{nej1,nej2} as
\begin{equation}\label{e57}
\mathcal{G}=\frac{1}{2}d\phi ^{i}H_{i}+e^{ \frac{1}{2}\alpha
_{i}\phi ^{i}}\:\mathbf{\Omega
}^{\alpha}_{\beta}\:d\chi^{\beta}E_{\alpha}.
\end{equation}
Now an affine or a Kozsul connection
\cite{hel,darling,tucker,thring,nakahara} on $G/K$ is a rule which
assigns to each $X\in E^{1}(G/K)$ a linear map
\begin{equation}\label{e58}
\nabla_{X}:E^{1}(G/K)\longrightarrow E^{1}(G/K),
\end{equation}
 such that
\begin{subequations}\label{e59}
\begin{align}
\nabla_{(fX+gY)}&=f\nabla_{X}+g\nabla_{Y},\notag\\
\notag\\
\nabla_{X}(fY)&=f(\nabla_{X}Y)+X(f)Y,\tag{\ref{e59}}
\end{align}
\end{subequations}
$\forall f,g\in C^{\infty}(G/K)$ and $\,\forall X,Y\in
E^{1}(G/K)$. Above $E^{1}(G/K)$ denotes the globally existing
module of the vector fields on $G/K$ which is composed of the
sections of the tangent bundle $TM(G/K)$. Now locally if we
consider the dual moving frame of \eqref{e54} namely $\{b_{c}\}$
such that
\begin{equation}\label{e60}
(e^{a},b_{c})=\delta^{a}_{\:\: c},
\end{equation}
where $\{b_{c}\}$ form up a local basis for $E^{1}(G/K)$ then we
can define the unique covariant exterior derivative $D$
\cite{hel,darling,tucker,thring,nakahara} associated with
\eqref{e58} as
\begin{equation}\label{e61}
D:\:\: S_{p}(TM(G/K))\longrightarrow S_{p+1}(TM(G/K)),
\end{equation}
where
\begin{equation}\label{e62}
S_{p}(TM(G/K))\cong E^{1}(G/K)\otimes E_{p}(G/K),
\end{equation}
with $E_{p}(G/K)$ denoting the module of $p$-forms on $G/K$. Now
if we consider the action of $D$ on the local moving frame
$\{b_{c}\}$ we have
\begin{equation}\label{e63}
Db_{a}=b_{c}\otimes\omega^{c}_{\; a},
\end{equation}
where the one-forms $\{\omega^{c}_{\; a}\}$ are called the
connection one-forms. Also we have
\begin{equation}\label{e64}
D(Db_{a})=b_{c}\otimes R^{c}_{\; a},
\end{equation}
where the two-forms $\{R^{c}_{\; a}\}$ are called the Ricci
curvature two-forms. In fact both $\omega^{c}_{\; a}$ and
$R^{c}_{\; a}$ are the components of the tangent space
endomorphism-valued differential forms $\omega\in
S_{1}(L(TM(G/K)))$ and $R\in S_{2}(L(TM(G/K)))$ respectively with
\begin{subequations}\label{e65.5}
\begin{align}
S_{p}(L(TM(G/K)))&\cong L(E^{1}(G/K)\longrightarrow
E^{1}(G/K))\otimes E_{p}(G/K)\notag\\
&\cong E^{1}(G/K)\otimes E_{p}(G/K)\otimes
E_{1}(G/K)\tag{\ref{e65.5}},
\end{align}
\end{subequations}
which is a local module isomorphism. From this definition we have
\begin{equation}\label{e65.6}
\omega=b_{c}\otimes\omega^{c}_{\; a}\otimes e^{a}\quad ,\quad
R=b_{c}\otimes R^{c}_{\; a}\otimes e^{a}.
\end{equation}
One can show that the components of these two objects satisfy
\cite{hel,darling,tucker,thring,nakahara}
\begin{equation}\label{e65}
R^{c}_{\; a}=d\omega^{c}_{\; a}+\omega^{c}_{\; d}\wedge
\omega^{d}_{\; a}.
\end{equation}
If one introduces the soldering form
\begin{equation}\label{e66}
\theta=b_{c}\otimes e^{c},
\end{equation}
then the torsion $T\in S_{2}(TM(G/K))$ can be defined as
\begin{equation}\label{e67}
T=D\theta=b_{c}\otimes(de^{c}+\omega^{c}_{\; d}\wedge e^{d}).
\end{equation}
Thus a torsion-free connection satisfies
\begin{equation}\label{e68}
de^{c}=-\omega^{c}_{\; d}\wedge e^{d}.
\end{equation}
For a $C^{\infty}$-manifold $M$, a pseudo-Riemannian or an
indefinite structure or metric on $M$ is a tensor field $g\in
T_{2}^{0}M$ such that $g(X,Y)=g(Y,X)$, $\,\forall X,Y\in T^{1}M$
and $\,\forall p\in M$, $g_{p}$ is a non-degenerate bilinear form
on $T_{p}M\times T_{p}M$. For a pseudo-Riemannian metric
$g(\cdot\,,\cdot)$ on $M$ an affine or a Kozsul connection
$\nabla$ on $TM(G/K)$ is said to be metric compatible if it
satisfies
\begin{equation}\label{e69}
X(g(Y,Z))=g(\nabla_{X}Y,Z)+g(Y,\nabla_{X}Z),
\end{equation}
$\forall X,Y,Z\in E^{1}(M)$. One can show that for a metric
compatible connection\footnote{For the following analysis in this
section we will raise or lower the indices by using the metric
components $g_{ab}$.}
\begin{equation}\label{e70}
dg_{ab}=\omega_{ab}+\omega_{ba}.
\end{equation}
Thus for a moving co-frame which generates constant metric
components likewise in \eqref{e54} the metric compatibility reads
\begin{equation}\label{e71}
\omega_{ab}=-\omega_{ba}.
\end{equation}
The fundamental theorem of Riemannian geometry
\cite{hel,darling,tucker,thring,nakahara} states that for a
pseudo-Riemannian structure $g(\cdot\,,\cdot)$ on a
$C^{\infty}$-manifold $M$ there exists a unique torsion-free,
metric compatible connection on the tangent bundle $TM(M)$ which
is called the Levi-Civita connection. Our point of view in this
section is to calculate the connection one-forms and the Ricci
curvature two-forms of the Levi-Civita connection associated with
the metric \eqref{e55} on $G/K$. The reason for the need of such
an explicit calculation within the brane dynamics we study in this
manuscript will be clear in the next two sections. Now if we
consider the moving co-frame \eqref{e54} following the review we
have done above which covers the basics of the pseudo-Riemannian
geometry, the connection one-forms of the Levi-Civita connection
of \eqref{e55} must satisfy the two Cartan structure equations
\begin{subequations}\label{e72}
\begin{align}
de^{c}&=-\omega^{c}_{\; d}\wedge e^{d},\notag\\
\notag\ \omega_{ab}&=-\omega_{ba}.\tag{\ref{e72}}
\end{align}
\end{subequations}
Owing to the gauge \eqref{e21} and the resulting axion-dilaton
coordinates we have specified for the bulk we will separate the
indices for the Cartan and the nilpotent directions as
\begin{equation}\label{e73}
\{a\}\longrightarrow\{i,\alpha\},
\end{equation}
where $a=1,...,$dim$(G/K)$ also $i=1,...,r$ and $\alpha=1,...,n$.
Their sum being $r+n=$dim$(G/K)$. Now in computing the connection
one-forms firstly after a direct calculation from \eqref{e54} we
find that
\begin{subequations}\label{e74}
\begin{align}
de^{i}&=0,\notag\\
\notag\ de^{\alpha}&=-\frac{1}{2}\alpha_{m}e^{\alpha
m}+C^{\alpha}_{\:\:\tau\rho}e^{\rho\tau}.\tag{\ref{e74}}
\end{align}
\end{subequations}
Here we define the coefficients
\begin{equation}\label{e75}
C^{\alpha}_{\:\:\tau\rho}=e^{\frac{1}{2}(\alpha_{j}-\tau_{j}-\rho_{j})\phi^{j}}D_{\gamma\:\:\:\beta}^{\:\:\alpha}
(\mathbf{\Omega}^{-1})^{\beta}_{\:\:\:\tau}(\mathbf{\Omega}^{-1})^{\gamma}_{\:\:\:\rho},
\end{equation}
where $D_{\gamma\:\:\:\beta}^{\:\:\alpha}\equiv
(\mathcal{D}_{\gamma})^{\alpha}_{\:\:\:\beta}$ are the components
of the $n\times n$ matrix
\begin{equation}\label{e76}
\mathcal{D}_{\gamma}=\frac{\partial\mathbf{\Omega}}{\partial\chi^{\gamma}},
\end{equation}
for which the definition and the computation is discussed in
detail in \cite{sssm1}. Since we have constant metric components
\begin{equation}\label{e77}
de_{a}=g_{ab}de^{b}.
\end{equation}
Thus lowering the indices in \eqref{e74} yields\footnote{The
reader should pay attention that if an index is free there is no
sum on it however we use the Einstein summation convention for all
the non-free indices.}
\begin{subequations}\label{e78}
\begin{align}
de_{i}&=-\frac{1}{2}g_{i\alpha}\alpha_{m}e^{\alpha
m}+g_{i\alpha}C^{\alpha}_{\:\:\tau\rho}e^{\rho\tau},\notag\\
\notag\ de_{\alpha}&=-\frac{1}{2}g_{\alpha\beta}\beta_{m}e^{\beta
m}+g_{\alpha\beta}C^{\beta}_{\:\:\tau\rho}e^{\rho\tau}.\tag{\ref{e78}}
\end{align}
\end{subequations}
For the viel-bein \eqref{e54} and its dual defined in \eqref{e60}
we have
\begin{subequations}\label{e79}
\begin{align}
(de^{a}\mid b_{c}\otimes b_{d})&=-(\omega^{a}_{\; b}\wedge
e^{b}\mid b_{c}\otimes b_{d})\notag\\
&=-(\omega^{a}_{\; d}\mid b_{c})+(\omega^{a}_{\; c}\mid
b_{d}),\tag{\ref{e79}}
\end{align}
\end{subequations}
where the parentheses define the action of the differential-forms
on the tensor fields and where we have made use of the
torsion-free equation from \eqref{e72} and the definition of the
wedge product
\begin{equation}\label{e80}
A\wedge B=A\otimes B-B\otimes A.
\end{equation}
By using \eqref{e79} together with the metric compatibility
equation from \eqref{e72} one can show that
\begin{equation}\label{e81}
(\omega_{ab}\mid b_{c})=\frac{1}{2}\big[(de_{b}\mid b_{c}\otimes
b_{a})+(de_{a}\mid b_{b}\otimes b_{c}) -(de_{c}\mid b_{a}\otimes
b_{b})\big],
\end{equation}
where we have also referred to \eqref{e77}. Now due to the
definition of the dual moving frame in \eqref{e60} one can
construct the connection one-forms which are introduced in
\eqref{e63} as
\begin{equation}\label{e82}
\omega_{ab}=(\omega_{ab}\mid b_{c})e^{c}.
\end{equation}
The components in the above expression can be calculated by direct
insertion of \eqref{e78} in \eqref{e81}. Doing so and bearing in
mind our index separation convention after a straightforward
calculation we find the following connection one-forms
\begin{subequations}\label{e83}
\begin{align}
\omega_{kj}=&\frac{1}{4}\;(g_{k\gamma}\gamma_{j}-g_{j\gamma}\gamma_{k})e^{\gamma},\notag\\
\notag\\
 \omega_{\gamma
j}=&\frac{1}{4}\;(g_{j\gamma}\gamma_{i}+g_{i\gamma}\gamma_{j})e^{i}+
\frac{1}{2}\;\big[\;g_{j\alpha}(C^{\alpha}_{\gamma\mu}-C^{\alpha}_{\mu\gamma})
+\frac{1}{2}\;g_{\gamma\mu}(\mu_{j}+\gamma_{j})\;\big]e^{\mu},\notag\\
\notag\\
\omega_{\gamma\sigma}=&\frac{1}{2}\;\big[\;\frac{1}{2}\;g_{\sigma\gamma}(\gamma_{i}-\sigma_{i})-
g_{i\alpha}(C^{\alpha}_{\sigma\gamma}-C^{\alpha}_{\gamma\sigma})\;\big]e^{i}\notag\\
\quad&+\frac{1}{2}\;\big[\;g_{\sigma\beta}(C^{\beta}_{\gamma\mu}-C^{\beta}_{\mu\gamma})+
g_{\gamma\beta}(C^{\beta}_{\mu\sigma}-C^{\beta}_{\sigma\mu})
+g_{\mu\beta}(C^{\beta}_{\gamma\sigma}-C^{\beta}_{\sigma\gamma})\;\big]e^{\mu}.
 \tag{\ref{e83}}
\end{align}
\end{subequations}
By directly plugging into the Cartan structure equations
\eqref{e72} the reader may verify that the one-forms in
\eqref{e83} are the correct ones satisfying
\eqref{e72}\footnote{We have indeed checked this, the antisymmetry
of $\omega_{kj},\omega_{\gamma\sigma}$ is straightforward, for
$\omega_{\gamma j}$ one also has to calculate $\omega_{j\gamma}$,
on the other hand for the torsion-free condition one has to
construct $-\omega_{cd}\wedge e^{d}$ after a moderately lengthy
computation for both species of free indices and then one can
prove that the results are equal to the expressions in
\eqref{e78}.}. If one lowers the free index in \eqref{e65} due to
the constancy of the metric components one gets
\begin{equation}\label{e84}
R_{ca}=d\omega_{ca}+\omega_{cd}\wedge \omega^{d}_{\; a}.
\end{equation}
Once the connection one-forms are calculated in \eqref{e83} it is
a straightforward but a handy task to compute the curvature
two-forms from \eqref{e84} with also the help of \eqref{e78}. We
should again refer to the index convention of \eqref{e54}. To
calculate the curvature two-forms in \eqref{e84} for three
different couples of index generations we will take the exterior
derivative of the connection one-forms in \eqref{e83}, then
compute the appropriate wedge products in \eqref{e84}, then add
the two, finally collect the factors within the two-form basis
generated by \eqref{e54}. This calculation for the Cartan
direction components of $R$ yields
\begin{subequations}\label{e85}
\begin{align}
R_{kj}=&\:\:\:\:\bigg(-\frac{1}{8}\gamma_{m}(g_{k\gamma}\gamma_{j}-g_{j\gamma}\gamma_{k})+
\frac{g^{i\alpha}}{16}(g_{k\gamma}\gamma_{i}-g_{i\gamma}\gamma_{k})(g_{j\alpha}\alpha_{m}+g_{m\alpha}\alpha_{j})\notag\\
 &\quad\:\:\:+\frac{g^{\alpha
n}}{16}(g_{k\alpha}\alpha_{m}+g_{m\alpha}\alpha_{k})(g_{n\gamma}\gamma_{j}-g_{j\gamma}\gamma_{n})
+\frac{g^{\alpha\tau}}{8}(g_{k\alpha}\alpha_{m}+g_{m\alpha}\alpha_{k})\notag\\
&\quad\:\:\:\times(g_{j\sigma}C^{\sigma}_{\tau\gamma}-g_{j\sigma}C^{\sigma}_{\gamma\tau}
+\frac{1}{2}g_{\tau\gamma}\gamma_{j}+\frac{1}{2}g_{\gamma\tau}\tau_{j})-\frac{g^{\alpha
\tau}}{8}(g_{j\tau}\tau_{m}+g_{m\tau}\tau_{j})\notag\\
 &\quad\:\:\:\times
(g_{k\sigma}C^{\sigma}_{\alpha\gamma}-g_{k\sigma}C^{\sigma}_{\gamma\alpha}
+\frac{1}{2}g_{\gamma\alpha}\alpha_{k}+\frac{1}{2}g_{\alpha\gamma}\gamma_{k})\bigg)e^{\gamma
m}\notag\\
&+\bigg(\frac{C^{\gamma}_{\kappa\rho}}{4}(g_{k\gamma}\gamma_{j}-g_{j\gamma}\gamma_{k})+
\frac{g^{i
n}}{16}(g_{k\rho}\rho_{i}-g_{i\rho}\rho_{k})(g_{n\kappa}\kappa_{j}-g_{j\kappa}\kappa_{n})\notag\\
&\quad\:\:\:+\frac{g^{i\alpha}}{8}(g_{k\rho}\rho_{i}-g_{i\rho}\rho_{k})
(g_{j\tau}C^{\tau}_{\alpha\kappa}-g_{j\tau}C^{\tau}_{\kappa\alpha}
+\frac{1}{2}g_{\alpha\kappa}\kappa_{j}+\frac{1}{2}g_{\kappa\alpha}\alpha_{j})\notag\\
 &\quad\:\:\:-\frac{g^{\alpha
n}}{8}(g_{n\kappa}\kappa_{j}-g_{j\kappa}\kappa_{n})
(g_{k\tau}C^{\tau}_{\alpha\rho}-g_{k\tau}C^{\tau}_{\rho\alpha}
+\frac{1}{2}g_{\alpha\rho}\rho_{k}+\frac{1}{2}g_{\rho\alpha}\alpha_{k})\notag\\
&\quad\:\:\:-\frac{g^{\alpha\tau}}{4}(g_{k\theta}C^{\theta}_{\alpha\rho}-g_{k\theta}C^{\theta}_{\rho\alpha}
+\frac{1}{2}g_{\alpha\rho}\rho_{k}+\frac{1}{2}g_{\rho\alpha}\alpha_{k})
(g_{j\sigma}C^{\sigma}_{\tau\kappa}-g_{j\sigma}C^{\sigma}_{\kappa\tau}\notag\\
&\quad\:\:\:+\frac{1}{2}g_{\tau\kappa}\kappa_{j}+\frac{1}{2}g_{\kappa\tau}\tau_{j})\bigg)e^{\rho\kappa}\notag\\
&-\bigg(\frac{g^{\alpha\tau}}{16}(g_{k\alpha}\alpha_{i}+g_{i\alpha}\alpha_{k})(g_{j\tau}\tau_{l}+g_{l\tau}\tau_{j})
\bigg)e^{il}. \tag{\ref{e85}}
\end{align}
\end{subequations}
On the other hand, since in nilpotent directions the expressions
in \eqref{e83} contain $C$-coefficients defined in \eqref{e75} we
have to consider their derivations. First of all we will define
\begin{equation}\label{e86}
d(D_{\gamma\:\:\:\beta}^{\:\:\alpha}
(\mathbf{\Omega}^{-1})^{\beta}_{\:\:\:\tau}(\mathbf{\Omega}^{-1})^{\gamma}_{\:\:\:\rho})
=Z^{\prime\alpha}_{\tau\rho\theta}e^{\theta},
\end{equation}
where
\begin{equation}\label{e87}
Z^{\prime\alpha}_{\tau\rho\theta}=U^{\alpha}_{\tau\rho\theta}+V^{\alpha}_{\tau\rho\theta}
+Y^{\alpha}_{\rho\tau\theta},
\end{equation}
with
\begin{subequations}\label{e88}
\begin{align}
U^{\alpha}_{\tau\rho\theta}&=e^{-\frac{1}{2}\theta_{i}\phi{i}}E_{\gamma\kappa\;\;\beta}^{\;\;\;\;\alpha}
(\mathbf{\Omega}^{-1})^{\beta}_{\:\:\:\tau}(\mathbf{\Omega}^{-1})^{\gamma}_{\:\:\:\rho}
(\mathbf{\Omega}^{-1})^{\kappa}_{\:\:\:\theta},\notag\\
\notag\\
 V^{\alpha}_{\tau\rho\theta}&=e^{-\frac{1}{2}\theta_{i}\phi{i}}D_{\gamma\:\:\:\beta}^{\:\:\alpha}
 F_{\xi\:\:\:\tau}^{\:\:\beta}
(\mathbf{\Omega}^{-1})^{\gamma}_{\:\:\:\rho}(\mathbf{\Omega}^{-1})^{\xi}_{\:\:\:\theta},\notag\\
\notag\\
Y^{\alpha}_{\rho\tau\theta}&=e^{-\frac{1}{2}\theta_{i}\phi{i}}D_{\gamma\:\:\:\beta}^{\:\:\alpha}
 F_{\xi\:\:\:\rho}^{\:\:\gamma}
(\mathbf{\Omega}^{-1})^{\beta}_{\:\:\:\tau}(\mathbf{\Omega}^{-1})^{\xi}_{\:\:\:\theta}.
 \tag{\ref{e88}}
\end{align}
\end{subequations}
Here we have also introduced the derivations
\begin{equation}\label{e89}
E_{\gamma\kappa\;\;\beta}^{\;\;\;\;\alpha}=\frac{\partial(
D_{\gamma\:\:\:\beta}^{\:\:\alpha})}{\partial\chi^{\kappa}}\quad,\quad
F_{\xi\:\:\:\tau}^{\:\:\beta}=\frac{\partial(
(\mathbf{\Omega}^{-1})^{\beta}_{\:\:\:\tau})}{\partial\chi^{\xi}}.
\end{equation}
Now if we further define
\begin{equation}\label{e90}
Z^{\alpha}_{\tau\rho\theta}=e^{\frac{1}{2}(\alpha_{j}-\tau_{j}-\rho_{j})\phi^{j}}Z^{\prime\alpha}_{\tau\rho\theta},
\end{equation}
and
\begin{equation}\label{e91}
\overline{Z}^{\alpha}_{\tau\rho
n}=\frac{1}{2}(\alpha_{n}-\tau_{n}-\rho_{n})
e^{\frac{1}{2}(\alpha_{j}-\tau_{j}-\rho_{j})\phi^{j}}D_{\gamma\:\:\:\beta}^{\:\:\alpha}
(\mathbf{\Omega}^{-1})^{\beta}_{\:\:\:\tau}(\mathbf{\Omega}^{-1})^{\gamma}_{\:\:\:\rho},
\end{equation}
we can eventually express the exterior derivative of the
$C$-coefficients which play essential role in the derivation of
the curvature two-forms in non-Cartan directions from connection
one-forms via \eqref{e84} as
\begin{equation}\label{e92}
dC^{\alpha}_{\:\:\tau\rho}=\overline{Z}^{\alpha}_{\tau\rho
n}e^{n}+Z^{\alpha}_{\tau\rho\theta}e^{\theta}.
\end{equation}
Now using the definitions in this computational result and
following the method we have mentioned before one can
systematically derive the components of the bundle-valued
curvature form $R$ in the nilpotent and the mixed directions. This
rigorous derivation for the mixed directions yields
\begin{subequations}\label{e93}
\begin{align}
R_{\alpha
j}=&\:\:\:\:\bigg(-\frac{1}{4}\mu_{m}g_{j\tau}(C^{\tau}_{\alpha\mu}-C^{\tau}_{\mu\alpha})
-\frac{1}{8}\mu_{m}g_{\mu\alpha}(\mu_{j}+\alpha_{j})-\frac{1}{2}g_{j\tau}(\overline{Z}^{\tau}_{\alpha\mu
m}-\overline{Z}^{\tau}_{\mu\alpha m})\notag\\
&\quad\:\:\:-\frac{1}{4}(g_{i\alpha}\alpha_{m}+g_{m\alpha}\alpha_{i})\big(\frac{g^{ik}}{4}
(g_{k\mu}\mu_{j}-g_{j\mu}\mu_{k})+\frac{g^{i\beta}}{2}(g_{j\sigma}C^{\sigma}_{\beta\mu}-g_{j\sigma}C^{\sigma}_{\mu\beta}
\notag\\
&\quad\:\:\:+\frac{1}{2}g_{\beta\mu}\mu_{j}+\frac{1}{2}g_{\mu\beta}\beta_{j})\big)
+\frac{g^{i\beta}}{8}(g_{i\tau}C^{\tau}_{\alpha\mu}-g_{i\tau}C^{\tau}_{\mu\alpha}
+\frac{1}{2}g_{\alpha\mu}\mu_{i}+\frac{1}{2}g_{\mu\alpha}\alpha_{i})\notag\\
&\quad\:\:\:\times(g_{j\beta}\beta_{m}+g_{m\beta}\beta_{j})
-\frac{1}{2}\big(\frac{g_{\beta\alpha}}{2}(\alpha_{m}-\beta_{m})
-g_{m\tau}(C^{\tau}_{\beta\alpha}-C^{\tau}_{\alpha\beta})\big)\notag\\
&\quad\:\:\:\times\big(\frac{g^{\beta k}}{4}
(g_{k\mu}\mu_{j}-g_{j\mu}\mu_{k})+\frac{g^{\beta\sigma}}{2}(g_{j\kappa}C^{\kappa}_{\sigma\mu}-g_{j\kappa}
C^{\kappa}_{\mu\sigma}+\frac{1}{2}g_{\sigma\mu}\mu_{j}\notag\\
&\quad\:\:\:+\frac{1}{2}g_{\mu\sigma}\sigma_{j})\big)
+\frac{g^{\beta\sigma}}{8}\big(g_{\beta\tau}(C^{\tau}_{\alpha\mu}-C^{\tau}_{\mu\alpha})+
g_{\alpha\tau}(C^{\tau}_{\mu\beta}-C^{\tau}_{\beta\mu})\notag\\
&\quad\:\:\:+g_{\mu\tau}(C^{\tau}_{\alpha\beta}-C^{\tau}_{\beta\alpha})\big)(g_{j\sigma}\sigma_{m}+g_{m\sigma}\sigma_{j})
\bigg)e^{\mu m}\notag\\
&+\bigg(\frac{g_{j\kappa}}{2}(Z^{\kappa}_{\alpha\tau
\rho}-Z^{\kappa}_{\tau\alpha
\rho})+\frac{1}{2}g_{j\kappa}C^{\mu}_{\tau\rho}(C^{\kappa}_{\alpha\mu}-C^{\kappa}_{\mu\alpha})
+\frac{1}{4}g_{\mu\alpha}C^{\mu}_{\tau\rho}(\mu_{j}+\alpha_{j})\notag\\
&\quad\:\:\:+\frac{1}{2}(g_{i\kappa}C^{\kappa}_{\alpha\rho}-g_{i\kappa}
C^{\kappa}_{\rho\alpha}+\frac{1}{2}g_{\alpha\rho}\rho_{i}+\frac{1}{2}g_{\rho\alpha}\alpha_{i})
\big(\frac{g^{ik}}{4}
(g_{k\tau}\tau_{j}-g_{j\tau}\tau_{k})\notag\\
&\quad\:\:\:+\frac{g^{i\beta}}{2}(g_{j\sigma}C^{\sigma}_{\beta\tau}-g_{j\sigma}
C^{\sigma}_{\tau\beta}+\frac{1}{2}g_{\beta\tau}\tau_{j}+\frac{1}{2}g_{\tau\beta}\beta_{j})\big)
+\frac{1}{2}\big(g_{\beta\kappa}(C^{\kappa}_{\alpha\rho}-C^{\kappa}_{\rho\alpha})\notag\\
&\quad\:\:\:+g_{\alpha\kappa}(C^{\kappa}_{\rho\beta}-C^{\kappa}_{\beta\rho})
+g_{\rho\kappa}(C^{\kappa}_{\alpha\beta}-C^{\kappa}_{\beta\alpha})\big)\big(\frac{g^{\beta
k}}{4} (g_{k\tau}\tau_{j}-g_{j\tau}\tau_{k})\notag\\
&\quad\:\:\:+\frac{g^{\beta\sigma}}{2}(g_{j\theta}C^{\theta}_{\sigma\tau}-g_{j\theta}
C^{\theta}_{\tau\sigma}+\frac{1}{2}g_{\sigma\tau}\tau_{j}+\frac{1}{2}g_{\tau\sigma}\sigma_{j})\big)\bigg)e^{\rho\tau}\notag\\
&+\bigg(\frac{g^{i\beta}}{16}
(g_{i\alpha}\alpha_{m}+g_{m\alpha}\alpha_{i})(g_{j\beta}\beta_{l}+g_{l\beta}\beta_{j})+\frac{g^{\beta\sigma}}{8}\big(
\frac{g_{\beta\alpha}}{2} (\alpha_{m}-\beta_{m})\notag\\
&\quad\:\:\:-g_{m\tau}(C^{\tau}_{\beta\alpha}-C^{\tau}_{\alpha\beta})\big)(g_{j\sigma}\sigma_{l}+g_{l\sigma}\sigma_{j})
\bigg)e^{ml},
 \tag{\ref{e93}}
\end{align}
\end{subequations}
which implicitly contains the derivation coefficients of $C$-terms
introduced in \eqref{e92}. Furthermore following a longer
computation the curvature two-forms in pure nilpotent directions
can be derived as
\begin{subequations}\label{e94}
\begin{align}
R_{\gamma\beta}=&\:\:\:\:\bigg(-\frac{g_{m\theta}}{2}(Z^{\theta}_{\beta\gamma\mu}-Z^{\theta}_{\gamma\beta\mu})
-\frac{1}{4}\mu_{m}\big(g_{\beta\theta}(C^{\theta}_{\gamma\mu}-C^{\theta}_{\mu\gamma})
+g_{\gamma\theta}(C^{\theta}_{\mu\beta}-C^{\theta}_{\beta\mu})\notag\\
 &\quad\:\:\:+g_{\mu\theta}(C^{\theta}_{\gamma\beta}-C^{\theta}_{\beta\gamma})\big)
-\frac{1}{2}\big(g_{\beta\theta}(\overline{Z}^{\theta}_{\gamma\mu
m}-\overline{Z}^{\theta}_{\mu\gamma
m})+g_{\gamma\theta}(\overline{Z}^{\theta}_{\mu \beta
m}-\overline{Z}^{\theta}_{\beta\mu m})\notag\\
&\quad\:\:\:+g_{\mu\theta}(\overline{Z}^{\theta}_{\gamma\beta
m}-\overline{Z}^{\theta}_{\beta\gamma
m})\big)-\frac{1}{4}(g_{j\gamma}\gamma_{m}+g_{m\gamma}\gamma_{j})\big(-\frac{g^{jk}}{2}
(g_{k\theta}C^{\theta}_{\beta\mu}\notag\\
&\quad\:\:\:-g_{k\theta}C^{\theta}_{\mu\beta}+\frac{1}{2}g_{\beta\mu}\mu_{k}+\frac{1}{2}g_{\mu\beta}\beta_{k})
+\frac{g^{j\alpha}}{2}\big(g_{\beta\theta}(C^{\theta}_{\alpha\mu}-C^{\theta}_{\mu\alpha})\notag\\
&\quad\:\:\:+g_{\alpha\theta}(C^{\theta}_{\mu\beta}-C^{\theta}_{\beta\mu})
 +g_{\mu\theta}(C^{\theta}_{\alpha\beta}-C^{\theta}_{\beta\alpha})\big)\big)
+\frac{1}{2}\big(g_{j\sigma}(C^{\sigma}_{\gamma\mu}-C^{\sigma}_{\mu\gamma})\notag\\
&\quad\:\:\:+\frac{1}{2}g_{\mu\gamma}(\mu_{j}+\gamma_{j})\big)
\big(-\frac{g^{jk}}{4}(g_{k\beta}\beta_{m}+g_{m\beta}\beta_{k})
+\frac{g^{j\alpha}}{2}\big( \frac{g_{\beta\alpha}}{2}
(\alpha_{m}-\beta_{m})\notag\\
&\quad\:\:\:-g_{m\theta}(C^{\theta}_{\beta\alpha}-C^{\theta}_{\alpha\beta})\big)\big)
-\frac{1}{2}\big( \frac{g_{\sigma\gamma}}{2}
(\gamma_{m}-\sigma_{m})-g_{m\alpha}(C^{\alpha}_{\sigma\gamma}-C^{\alpha}_{\gamma\sigma})\big)\notag\\
&\quad\:\:\:\times\big(-\frac{g^{\sigma k}}{2}
(g_{k\theta}C^{\theta}_{\beta\mu}
-g_{k\theta}C^{\theta}_{\mu\beta}+\frac{1}{2}g_{\beta\mu}\mu_{k}+\frac{1}{2}g_{\mu\beta}\beta_{k})\notag\\
&\quad\:\:\:+\frac{g^{\sigma\kappa}}{2}\big(g_{\beta\theta}(C^{\theta}_{\kappa\mu}-C^{\theta}_{\mu\kappa})
+g_{\kappa\theta}(C^{\theta}_{\mu\beta}-C^{\theta}_{\beta\mu})
 +g_{\mu\theta}(C^{\theta}_{\kappa\beta}-C^{\theta}_{\beta\kappa})\big)\big)\notag\\
&\quad\:\:\:+\frac{1}{2}\big(g_{\sigma\alpha}(C^{\alpha}_{\gamma\mu}-C^{\alpha}_{\mu\gamma})
+g_{\gamma\alpha}(C^{\alpha}_{\mu\sigma}-C^{\alpha}_{\sigma\mu})
 +g_{\mu\alpha}(C^{\alpha}_{\gamma\sigma}-C^{\alpha}_{\sigma\gamma})\big)\notag\\
&\quad\:\:\:\times\big(-\frac{g^{\sigma
k}}{4}(g_{k\beta}\beta_{m}+g_{m\beta}\beta_{k})
+\frac{g^{\sigma\kappa}}{2}\big( \frac{g_{\beta\kappa}}{2}
(\kappa_{m}-\beta_{m})\notag\\
&\quad\:\:\:-g_{m\theta}(C^{\theta}_{\beta\kappa}-C^{\theta}_{\kappa\beta})\big)\big)\bigg)e^{\mu
m}\notag\\
&+\bigg(\frac{1}{2}\big(g_{\beta\theta}(Z^{\theta}_{\gamma\tau\rho}-Z^{\theta}_{\tau\gamma\rho})
+g_{\gamma\theta}(Z^{\theta}_{\tau\beta\rho}-Z^{\theta}_{\beta\tau\rho})
+g_{\tau\theta}(Z^{\theta}_{\gamma\beta\rho}-Z^{\theta}_{\beta\gamma\rho})\big)\notag\\
&\quad\:\:\:+\frac{C^{\mu}_{\tau\rho}}{2}\big(g_{\beta\theta}(C^{\theta}_{\gamma\mu}-C^{\theta}_{\mu\gamma})
+g_{\gamma\theta}(C^{\theta}_{\mu\beta}-C^{\theta}_{\beta\mu})
+g_{\mu\theta}(C^{\theta}_{\gamma\beta}-C^{\theta}_{\beta\gamma})\big)\notag\\
&\quad\:\:\:+\frac{1}{2}\big( \frac{g_{\rho\gamma}}{2}
(\rho_{j}+\gamma_{j})+g_{j\sigma}(C^{\sigma}_{\gamma\rho}-C^{\sigma}_{\rho\gamma})\big)
\big(-\frac{g^{jk}}{2} (g_{k\theta}C^{\theta}_{\beta\tau}\notag\\
&\quad\:\:\:-g_{k\theta}C^{\theta}_{\tau\beta}+\frac{1}{2}g_{\beta\tau}\tau_{k}+\frac{1}{2}g_{\tau\beta}\beta_{k})
+\frac{g^{j\alpha}}{2}\big(g_{\beta\theta}(C^{\theta}_{\alpha\tau}-C^{\theta}_{\tau\alpha})\notag\\
&\quad\:\:\:+g_{\alpha\theta}(C^{\theta}_{\tau\beta}-C^{\theta}_{\beta\tau})
 +g_{\tau\theta}(C^{\theta}_{\alpha\beta}-C^{\theta}_{\beta\alpha})\big)\big)
+\frac{1}{2}\big(g_{\sigma\alpha}(C^{\alpha}_{\gamma\rho}-C^{\alpha}_{\rho\gamma})\notag\\
&\quad\:\:\:+g_{\gamma\alpha}(C^{\alpha}_{\rho\sigma}-C^{\alpha}_{\sigma\rho})
 +g_{\rho\alpha}(C^{\alpha}_{\gamma\sigma}-C^{\alpha}_{\sigma\gamma})\big)
 \big(-\frac{g^{\sigma k}}{2} (g_{k\theta}C^{\theta}_{\beta\tau}\notag\\
&\quad\:\:\:-g_{k\theta}C^{\theta}_{\tau\beta}+\frac{1}{2}g_{\beta\tau}\tau_{k}+\frac{1}{2}g_{\tau\beta}\beta_{k})
+\frac{g^{\sigma\kappa}}{2}\big(g_{\beta\theta}(C^{\theta}_{\kappa\tau}-C^{\theta}_{\tau\kappa})\notag\\
&\quad\:\:\:+g_{\kappa\theta}(C^{\theta}_{\tau\beta}-C^{\theta}_{\beta\tau})
 +g_{\tau\theta}(C^{\theta}_{\kappa\beta}-C^{\theta}_{\beta\kappa})\big)\big)\bigg)e^{\rho\tau}\notag\\
 &+\bigg(-\frac{1}{2}g_{l\alpha}(\overline{Z}^{\alpha}_{\beta\gamma
m}-\overline{Z}^{\alpha}_{\gamma\beta m})
+\frac{1}{4}(g_{j\gamma}\gamma_{m}+g_{m\gamma}\gamma_{j})
\big(-\frac{g^{jk}}{4}(g_{k\beta}\beta_{l}\notag\\
&\quad\:\:\:+g_{l\beta}\beta_{k}) +\frac{g^{j\alpha}}{2}\big(
\frac{g_{\beta\alpha}}{2}
(\alpha_{l}-\beta_{l})-g_{l\theta}(C^{\theta}_{\beta\alpha}-C^{\theta}_{\alpha\beta})\big)\big)\notag\\
&\quad\:\:\:+\frac{1}{2}\big( \frac{g_{\sigma\gamma}}{2}
(\gamma_{m}-\sigma_{m})-g_{m\alpha}(C^{\alpha}_{\sigma\gamma}-C^{\alpha}_{\gamma\sigma})\big)
\big(-\frac{g^{\sigma
k}}{4}(g_{k\beta}\beta_{l}+g_{l\beta}\beta_{k})\notag\\
&\quad\:\:\:+\frac{g^{\sigma\kappa}}{2}\big(
\frac{g_{\beta\kappa}}{2} (\kappa_{l}-\beta_{l})
-g_{l\theta}(C^{\theta}_{\beta\kappa}-C^{\theta}_{\kappa\beta})\big)\big)\bigg)e^{ml}.
\tag{\ref{e94}}
\end{align}
\end{subequations}
By carefully inspecting \eqref{e85}, \eqref{e93}, and \eqref{e94}
one can verify that they satisfy
\begin{equation}\label{e95}
R_{kj}=-R_{jk}\quad ,\quad R_{\alpha j}=-R_{j\alpha}\quad ,\quad
R_{\gamma\beta}=-R_{\beta\gamma},
\end{equation}
which is a direct consequence of the metric compatibility
\eqref{e71} when used in \eqref{e84}\footnote{Again to show the
second identity in \eqref{e95} one has to calculate $R_{j\alpha}$
separately.}.
\section{Gauss Equation}
In this section, we will focus on the details of the geometry of
the braneworld immersion in the bulk. In Section two, we have
discussed that the metric \eqref{e2} on the braneworld is induced
by the bulk metric. In the Nambu-Goto action this fact is
introduced as a definition so that the action defines a multiple
of the pseudo-Riemannian volume of the braneworld in the bulk.
Whereas for the equivalent Polyakov action \eqref{e2} appears as a
field equation of the braneworld metric which is made a dynamic
field. Both of these approaches require that the braneworld and
the bulk metrics are related by the push forward map of the
inclusion map of the braneworld in the bulk. In this manuscript
our main framework are the cases when the braneworld is an
immersion \cite{darling} in the bulk so that the inclusion map is
a $C^{\infty}$-map and its differential map is injective. If the
immersion itself is injective then we will have an imbedding. A
more restrictive case would be assuming the braneworld to be a
submanifold of the bulk which would bring more restrictions on the
submersion characteristics of the differential (atlas) structures
of the braneworld and the bulk \cite{darling}. Therefore by
requiring the relation \eqref{e2} we geometrize the
pseudo-Riemannian braneworld as an isometric immersion in the
pseudo-Riemannian bulk. In this case \eqref{e2} becomes the first
fundamental form. Since the Nambu-Goto action is proportional to
the volume of the braneworld in the bulk the variational principle
leads to the minimality condition thus the braneworld becomes a
minimal immersion or submanifold of the bulk. Our formulation in
this section will follow the outlines of
\cite{aminov,docarmo,sternberg,oneil}. We should first observe
that
\begin{equation}\label{e96}
\{\frac{\partial}{\partial\varphi^{1}},\cdots
,\frac{\partial}{\partial\varphi^{r}},
\frac{\partial}{\partial\varphi^{r+1}},\cdots
,\frac{\partial}{\partial\varphi^{r+n}}\}=\{
\frac{\partial}{\partial\phi^{1}},\cdots
,\frac{\partial}{\partial\phi^{r}},\frac{\partial}{\partial\chi^{1}},\cdots
,\frac{\partial}{\partial\chi^{n}}\},
\end{equation}
is a local coordinate frame on the bulk\footnote{The reader should
take $\partial/\partial\varphi^{a}$ as the push-forward of the
coordinate basis frame on $\Bbb{R}^{(r+n)}$ via the inverse of the
coordinate chart map.} the dual bulk moving co-frame being
$\{d\varphi^{a}\}=\{d\phi^{i},d\chi^{\alpha}\}$. By bearing in
mind that on the braneworld $\varphi^{a}=\varphi^{a}(x^{1},\cdots
,x^{p+1})$, for $A=1,\cdots ,p+1$ if one considers the set
\begin{equation}\label{e97}
\{r_{A}\}=\{\partial_{A}\varphi^{a}\frac{\partial}{\partial\varphi^{a}}\},
\end{equation}
of vector fields at the bulk-braneworld intersection then their
restriction to the world volume becomes a moving frame on the
braneworld \cite{aminov}. At this stage we will introduce the
orthogonal tangent space decomposition of the bulk at any point
$p\in W \subset G/K$
\begin{equation}\label{e98}
T_{p}(G/K)=T_{p}(W)\oplus \big [T_{p}(W)\big ]^{\bot},
\end{equation}
where if $X\in T_{p}(W)$ and $Y\in [T_{p}(W)\big ]^{\bot}$ then
\begin{equation}\label{e99}
g(X,Y)=0.
\end{equation}
This decomposition induces a local module decomposition of vector
fields on the bulk\footnote{In fact, meaningfully bulk vector
fields on the intersection of the bulk and the braneworld.}
\begin{equation}\label{e100}
E^{1}(G/K)=H(W)\oplus V(W),
\end{equation}
with horizontal vector fields $H(W)$ such that $H(W)\cong
E^{1}(W)$ and their vertical complements $V(W)$\footnote{This
terminology switches for the submersions.}. Therefore \eqref{e97}
becomes a local basis for $H(W)$. For our construction of the
Gauss equation which relates the curvature elements of the bulk to
the braneworld in addition to the coordinate frame \eqref{e96} and
the co-frame \eqref{e54} which we made use of in the previous
section we will consider a third frame for the bulk based on the
decomposition \eqref{e98} namely
\begin{equation}\label{e101}
\{e^{\prime}_{a}\}=\{r_{A},\xi_{m}\},
\end{equation}
where $m=1,\cdots ,r+n-(p+1)$, and $\{\xi_{m}\}$ is an arbitrary
local frame for the vertical vector fields $V(W)$. Since
$\varphi^{a}=\varphi^{a}(x^{A})$ on the world volume we have
\begin{equation}\label{e101.5}
d\varphi^{a}=\partial_{A}\varphi^{a}dx^{A}.
\end{equation}
If now we consider the moving co-frame
$\{\tilde{e}^{\:a}\}=\{dx^{A},\tilde{\xi}^{\:m}\}$ on the bulk we
have the transformation
\begin{equation}\label{e101.6}
d\varphi^{a}=S^{a}_{\:\:\:b}\tilde{e}^{\:b}=\partial_{A}\varphi^{a}dx^{A}+S^{a}_{\:\:\:m}
\tilde{\xi}^{\:m},
\end{equation}
which gives \eqref{e101.5} when restricted onto the braneworld.
From \eqref{e101.6} we see that we have the dual moving frame
transformation
\begin{equation}\label{e101.7}
\tilde{b}_{a}=\frac{\partial}{\partial\varphi^{b}}S^{b}_{\:\:\:a},
\end{equation}
where $\{\tilde{b}_{a}\}=\{\frac{\partial}{\partial
x^{A}},\xi_{m}\}$. When we compare \eqref{e97} with \eqref{e101.7}
we conclude that
\begin{equation}\label{e101.8}
r_{A}=\frac{\partial}{\partial x^{A}},
\end{equation}
which justifies that \eqref{e97} is a moving frame for the
braneworld $W$. Now let us introduce the orthogonal projection
operators
\begin{equation}\label{e102}
\mathcal{H}\: :\: E^{1}(G/K)\longrightarrow H(W)\quad ,\quad
\mathcal{V}\: :\: E^{1}(G/K)\longrightarrow V(W).
\end{equation}
We can introduce the $\mathcal{T}$-tensor by first defining
\begin{equation}\label{e102}
\mathcal{T}_{E}\:
F=\mathcal{V}(\nabla_{\mathcal{H}E}(\mathcal{H}F))+
\mathcal{H}(\nabla_{\mathcal{H}E}(\mathcal{V}F)),
\end{equation}
where $E,F\in E^{1}(G/K)$. Here $\nabla$ is the Levi-Civita
connection of the bulk metric \eqref{e55} on the bulk. Now if we
define
\begin{equation}\label{e102.5}
\mathcal{T}(X_{1},X_{2},\omega)=<\mathcal{T}_{E}\: F,\omega>,
\end{equation}
then $\mathcal{T}\in T^{1}_{\:\:\:2}(G/K)$\footnote{With the help
of the bulk metric \eqref{e102} can also be used to define a
$T_{3}(G/K)$ tensor.}. When $X_{1},X_{2}\in H(W)$ then the second
term in \eqref{e102} drops and we are led to the definition of the
second fundamental form or the extrinsic curvature \cite{wheeler}
of the braneworld immersion
\begin{equation}\label{e103}
B(X_{1},X_{2})=\mathcal{V}(\nabla_{\mathcal{H}X_{1}}(\mathcal{H}X_{2}))
=\mathcal{V}(\nabla_{X_{1}}(X_{2}))\equiv
(\nabla_{X_{1}}(X_{2}))^{\bot},
\end{equation}
whose image is clearly an element of $V(W)$. One can show that
\begin{equation}\label{e104}
B(X_{1},X_{2})=\nabla_{X_{1}}(X_{2})-\nabla^{W}_{X_{1}}(X_{2}),
\end{equation}
where $\nabla^{W}$ is the Levi-Civita connection on the braneworld
corresponding to the first-fundamental form or the induced metric
defined in \eqref{e2}. We should state that for the second term in
\eqref{e104} which is an element of $H(W)$ we consider the
restrictions of the bulk vector fields $X_{1},X_{2}\in H(W)$ on
the braneworld which are full copies of the formers. Comparing
\eqref{e103} and \eqref{e104} we see that
\begin{equation}\label{e105}
\nabla^{W}_{X_{1}}(X_{2})=\mathcal{H}(\nabla_{X_{1}}(X_{2})).
\end{equation}
Since $\mathcal{T}\in T^{1}_{\:\:\:2}(G/K)$ and since
$B(X_{1},X_{2})\in V(W)$ by using the bulk frame \eqref{e101} we
can introduce the components of the second fundamental form as
\begin{equation}\label{e106}
B=L_{AB}^{m}\:\tilde{r}^{A}\otimes\tilde{r}^{B}\otimes\xi_{m}.
\end{equation}
Here $\tilde{r}^{A},\tilde{r}^{B}$ are the dual one-forms of
$r_{A},r_{B}$. We prefer using the notation of \cite{aminov}. One
can keep also the $\{r_{A}\}$ elements of \eqref{e101} arbitrary
likewise the vertical complements instead of specifying them in
\eqref{e97} and then define the second fundamental form
components. However the practical essence of the frame
\eqref{e101} will show itself when we write the Gauss equation
since it contains the $\{r_{A}\}$ part as a frame for the
braneworld. Now if $X_{1}=x_{1}^{A}r_{A}$ and
$X_{2}=x_{2}^{B}r_{B}$ then we have
\begin{equation}\label{e107}
B(X_{1},X_{2})=L_{AB}^{m}\, x_{1}^{A}\, x_{2}^{B}\,\xi_{m}.
\end{equation}
Also
\begin{equation}\label{e108}
B(r_{A},r_{B})=L_{AB}^{m}\,\xi_{m}.
\end{equation}
Thus from \eqref{e104} we have
\begin{subequations}\label{e109}
\begin{align}
\nabla_{r_{A}}(r_{B})&=\nabla^{W}_{r_{A}}(r_{B})+L_{AB}^{m}\,\xi_{m}\notag\\
\notag\\
&=\tilde{\Gamma}_{AB}^{\:C}r_{C}+L_{AB}^{m}\,\xi_{m}\tag{\ref{e109}},
\end{align}
\end{subequations}
where $\tilde{\Gamma}_{AB}^{\:C}$ are the Christoffel symbols with
respect to the basis \eqref{e97} of the induced metric on the
braneworld. By using \eqref{e99} from \eqref{e109} we can derive
that
\begin{equation}\label{e110}
g(\nabla_{r_{A}}(r_{B}),\xi_{n})=L_{AB}^{m}g_{mn},
\end{equation}
where $g_{mn}=g(\xi_{m},\xi_{n})$. At this point if we switch back
to the coordinate basis on the bulk the connection entry above can
be calculated as \cite{aminov,docarmo}
\begin{subequations}\label{e111}
\begin{align}
\nabla_{r_{A}}(r_{B})=\nabla_{\partial_{A}\varphi^{a}\frac{\partial}{\partial\varphi^{a}}}
(\partial_{B}\varphi^{b}\frac{\partial}{\partial\varphi^{b}})&=
\big
[\partial_{A}\varphi^{b}\frac{\partial(\partial_{B}\varphi^{c})}{\partial\varphi^{b}}
+\partial_{A}\varphi^{a}\partial_{B}\varphi^{b}\Gamma_{ab}^{c}\big]\:\frac{\partial}
{\partial\varphi^{c}}\notag\\
\notag\\
&=\big [\partial^{2}_{AB}\varphi^{c}
+\partial_{A}\varphi^{a}\partial_{B}\varphi^{b}\Gamma_{ab}^{c}\big]\:\frac{\partial}
{\partial\varphi^{c}}\tag{\ref{e111}},
\end{align}
\end{subequations}
where we define
\begin{equation}\label{e112}
\partial^{2}_{AB}\varphi^{c}=\frac{\partial^{2}\varphi^{c}}{\partial
x^{A}\partial x^{B}}.
\end{equation}
In \eqref{e111} the Cristoffel symbols belong to the bulk metric
and they are with respect to the coordinate frame \eqref{e96}
namely
\begin{equation}\label{e113}
\nabla_{\frac{\partial} {\partial\varphi^{a}}}(\frac{\partial}
{\partial\varphi^{b}})=\Gamma^{c}_{ab}\frac{\partial}
{\partial\varphi^{c}}.
\end{equation}
One can calculate them from the bulk metric components in
\eqref{e40} as \cite{tucker}
\begin{equation}\label{e114}
\Gamma_{ab}^{c}=\frac{1}{2}g^{cd}\big[\partial_{a}(g_{bd})
+\partial_{b}(g_{da})-\partial_{d}(g_{ab})\big],
\end{equation}
where we omit the torsion terms since we calculate the Levi-Civita
connection coefficients, as well as the basis structure constant
terms since our basis is a coordinate one. In \eqref{e114} we use
$\partial_{a}=\partial/\partial\varphi^{a}$. Now the second
fundamental form coefficients in \eqref{e110} can be expressed in
terms of the bulk coordinates as
\begin{equation}\label{e115}
L_{AB}^{m}g_{mn}=\mathcal{F}_{ABn}(\varphi^{a},\partial\varphi^{a},\partial^{2}\varphi^{a}),
\end{equation}
where the functional $\mathcal{F}_{ABn}$ is
\begin{equation}\label{e116}
\mathcal{F}_{ABn}(\varphi^{a},\partial\varphi^{a},\partial^{2}\varphi^{a})=\big
[\partial^{2}_{AB}\varphi^{c}
+\partial_{A}\varphi^{a}\partial_{B}\varphi^{b}\Gamma_{ab}^{c}\big]\xi_{n}^{e}g_{ce},
\end{equation}
where we have defined the components of the vertical frame via
$\xi_{n}=\xi_{n}^{e}\frac{\partial}{\partial\varphi^{e}}$ which
are subject to the conditions (which can be read in terms of the
bulk coordinates from)
\begin{equation}\label{e116.5}
\partial_{A}\varphi^{a}\xi_{m}^{b}g_{ab}=0,
\end{equation}
resulting from \eqref{e99} and for which $A=1,\cdots ,p+1$ and
$m=1,\cdots ,n+r-(p+1)$. We should state that since the second
fundamental form is symmetric $B(X_{1},X_{2})=B(X_{2},X_{1})$ we
have $L_{AB}^{m}=L_{BA}^{m}$ which can easily be seen from
\eqref{e108}. In addition to the above mentioned method on the
other hand if one calculates the metric components and the
corresponding Christoffel symbols $\Gamma^{\,\prime
\,c}_{\:\:\:ab}$ with respect to the basis \eqref{e101} then with
the help of \eqref{e99} from \eqref{e110} one can directly find
the second fundamental form components as
\begin{equation}\label{e117}
L_{AB}^{m}=\Gamma^{\,\prime \,m}_{\:\:\:AB}.
\end{equation}
Having defined the second fundamental form, to construct the Gauss
equation we will now consider the metric dual of the bulk Riemann
tensor whose action can be obtained via
$R(X,Y,Z,T)=g(R(X,Y)\,Z,T)$ for $X,Y,Z,T\in E^{1}(G/K)$. Its
components in a bulk moving frame $\{b_{a}\}$ can be found as
$R_{abcd}=g(R(b_{a},b_{b})\,b_{c},b_{d})$. Here we have introduced
\begin{equation}\label{e118}
R(X,Y)=\nabla_{X}\nabla_{Y}-\nabla_{Y}\nabla_{X}-\nabla_{[X,Y]}.
\end{equation}
If $n=0,\cdots,4$ denotes the number of horizontal vector fields
in the set $X,Y,Z,T\in E^{1}(G/K)$ from the computation of
$R(X,Y,Z,T)$ one can obtain the five fundamental equations of the
braneworld immersion in the bulk including the Gauss ($n=4$),
Ricci ($n=2$), and the Codazzi ($n=3$) equations
\cite{aminov,docarmo,sternberg,oneil}. We will focus only on the
Gauss equation for our main objective of relating the bulk and the
braneworld curvatures. It reads
\begin{subequations}\label{e119}
\begin{align}
g(R(X,Y)\,Z,T)&=G(R^{W}(X,Y)\,Z,T)+g\big
(B(X,T),B(Y,Z)\big)\notag\\
\notag\\
&\:\:\:\:-g\big (B(Y,T),B(X,Z)\big)\tag{\ref{e119}},
\end{align}
\end{subequations}
where $G(R^{W}(X,Y)\,Z,T)$ is the action of the index lowered
Riemann tensor of the braneworld\footnote{Corresponding to the
Levi-Civita connection of the induced metric on the braneworld.}
on the horizontal vector fields $X,Y,Z,T$ which have exact copies
in $E^{1}(W)$. As we have stated before by using the special bulk
frame \eqref{e101} which contains a braneworld frame in it we can
use the Gauss equation \eqref{e119} to relate the Riemann tensor
components of the bulk and the braneworld. Direct substitution of
\eqref{e101} in \eqref{e119} gives
\begin{equation}\label{e120}
R^{W}_{ABCD}=R_{ABCD}+\big[L_{AC}^{m}L_{BD}^{n}-L_{BC}^{m}L_{AD}^{n}\big]g_{mn}.
\end{equation}
We may express this equation in terms of the functionals defined
in \eqref{e116} as
\begin{equation}\label{e121}
R^{W}_{ABCD}=R_{ABCD}+\mathcal{F}_{AC}^{\:\:\:\:\:\:m}\mathcal{F}_{BDm}
-\mathcal{F}_{BC}^{\:\:\:\:\:\:m}\mathcal{F}_{ADm},
\end{equation}
where $\mathcal{F}_{AC}^{\:\:\:\:\:\:m}=\mathcal{F}_{ACn}g^{nm}$
with $g^{nm}$ being the inverse of $g_{nm}=g(\xi_{n},\xi_{m})$.
\section{Gravitating Dynamic Branes}
In this section, we will consider the gravity sector of a dynamic
brane coupled to world volume gravity and matter fields that is
immersed in a symmetric space bulk.  The gravity and the matter
sectors will be coupled to the induced braneworld metric
\eqref{e2} which arise from the natural definition of the
braneworld dynamics where the braneworld is an isometric immersion
in the bulk. In Section two, we have considered the free brane
motion. When one applies the least action principle to the
Nambu-Goto action of \eqref{e7} the field equations one derives
correspond to a minimal isometric immersion since \eqref{e7} is
proportional to the volume of the braneworld in the bulk. These
field equations are equivalent to the vanishing of the vertical
mean curvature vector field \cite{aminov,docarmo,chern,lawson}
which is defined as
\begin{equation}\label{e121.5}
H=\frac{1}{p+1}L_{AB}^{m}G^{AB}\xi_{m}.
\end{equation}
Thus one may simply replace the field equations of \eqref{e7} with
$L_{A}^{m\;A}=L_{AB}^{m}G^{AB}=0$ for $m=1,\cdots, n+r-(p+1)$.
Therefore from the second fundamental form point of view the free
brane motion has well defined restrictions which relate it to the
theory of analytical functions \cite{chern,lawson}. However in
this section due to the presence of accompanying fields variation
of the total action will not result in a minimality condition for
the Nambu-Goto term (its variation is not equal to zero). This
means that the braneworld volume is not minimized any more and we
do not have simply a minimal isometric immersion still an
isometric immersion though. Thus the second fundamental form
geometry of the immersion, though restricted further has a more
complicated role in the overall dynamics. Bearing in mind this
fact our primary objective in this section will be to combine the
ingredients from the previous sections in order to find a
geometrical method of implementing into the Einstein equation the
induced braneworld metric condition which may be considered as a
constraint equation in an alternative approach which enables the
reduction of the brane action term naturally to a new cosmological
constant term in the overall action. By doing so one may derive
the Einstein equations by means of the usual independent metric
variation methods then one can express these equations solely in
terms of the bulk coordinates of the braneworld by implementing
the above mentioned constraint. In this manner, we may also be
able to express the gravity dynamics explicitly in terms of the
second fundamental form (the extrinsic curvature). Now as a start
let us consider the gravity and the matter coupled dynamic
braneworld action which can be given as
\begin{subequations}\label{e122}
\begin{align}
 S_{Brane}&=-T_{p}\, \int_{W} d^{(p+1)}\sigma\sqrt{-\text{det}(g_{ab}\partial_{A}
 \varphi^{a}\partial_{B}\varphi^{b})}\notag\\
 \notag\\
 &\:\:\:\:\:+\int_{W}\bigg(\;\frac{1}{16\kappa\pi}\;R_{AB}^{(W)}\wedge\ast e^{AB}+\ast\Lambda\bigg)
 +\int_{W}\mathcal{L}_{Matter},\tag{\ref{e122}}
 \end{align}
\end{subequations}
where $G_{AB}=g_{ab}\partial_{A}
 \varphi^{a}\partial_{B}\varphi^{b}$ is the induced metric on the braneworld from the bulk which
 originates from the physical kinematics of the free brane dynamics\footnote{In order that the free brane mimics a relativistic particle
 its world volume must be minimized so that its metric must be the induced one and it must be an isometric immersion.}.
 Here the Hodge star operator is defined with respect to
this induced braneworld metric \eqref{e2}. In the above action
$(W)$ stands for the braneworld fields. Also while the little
Latin indices correspond to the bulk the capital ones correspond
to the braneworld. We will take the moving co-frame on the
braneworld as the coordinate co-frame
$\{e^{A}\}=\{dx^{A}\}=\{\tilde{r}^{A}\}$ of the previous section.
Thus the volume-form $d^{(p+1)}\sigma$ is also constructed from
this co-frame and the braneworld metric components are calculated
with respect to it. Both in \eqref{e122} and in the following we
prefer to use the notation $e^{ABC\cdots}=e^{A}\wedge e^{B}\wedge
e^{C}\wedge\cdots$. In its most pure form in the above action
apart from the matter fields the braneworld coordinates
$\{\varphi^{a}(x^{A})\}$ (as scalar fields on the braneworld) are
the only independent fields. If the bulk does not admit any
dynamics or if the braneworld dynamics do not bring any
constraints on it then in the simplest case of \eqref{e122} one
may start with a bulk metric independently and derive the
expression for the induced braneworld metric. On the other hand in
the most general case if the bulk carries a dynamical structure on
it which may or may not be induced by the braneworld then
$G_{AB}=g_{ab}\partial_{A}
 \varphi^{a}\partial_{B}\varphi^{b}$ links the two dynamics. In either case not only the braneworld metric $G_{AB}=g_{ab}\partial_{A}
 \varphi^{a}\partial_{B}\varphi^{b}$ but also the related gravity structures become functionals of the braneworld coordinates. As the reader
may quickly realize in its pure form in which \eqref{e122}
 is considered solely in terms of $\{\varphi^{a}(x^{A})\}$ the derivation of the
variation of \eqref{e122} in terms of $\{\varphi^{a}(x^{A})\}$
becomes a highly non-standard and a cumbersome computation. This
is due to the fact that as a result of the induced metric
$G_{AB}=g_{ab}\partial_{A}
 \varphi^{a}\partial_{B}\varphi^{b}$
\begin{equation}\label{e122.5}
R_{AB}^{(W)}=R_{AB}^{(W)}(\varphi^{a}),\quad
\ast=\ast(\varphi^{a}).
\end{equation}
Due this complification we will follow a different track in which
we will replace \eqref{e122} with a constraint system. In
\eqref{e122} in addition to the matter fields and the braneworld
coordinates $\{\varphi^{a}(x^{A})\}$ we will take the braneworld
metric $G_{AB}$ as independent too. Therefore we see that in this
case \eqref{e2} namely $G_{AB}=g_{ab}\partial_{A}
 \varphi^{a}\partial_{B}\varphi^{b}$ relates these independent fields
to each other so it must be taken as a set of constraint equations
in this new approach. Thus we have the equivalent system
\begin{subequations}\label{e122.6}
\begin{align}
 S_{Brane}&=-T_{p}\, \int_{W} d^{(p+1)}\sigma\sqrt{-\text{det}(g_{ab}\partial_{A}
 \varphi^{a}\partial_{B}\varphi^{b})}\notag\\
 \notag\\
 &\:\:\:\:\:+\int_{W}\bigg(\;\frac{1}{16\kappa\pi}\;R_{AB}^{(W)}\wedge\ast e^{AB}+\ast\Lambda\bigg)
 +\int_{W}\mathcal{L}_{Matter}\notag\\
\notag\\
G_{AB}&=g_{ab}\partial_{A}
 \varphi^{a}\partial_{B}\varphi^{b}\tag{\ref{e122.6}},
 \end{align}
\end{subequations}
Now in our new system through the action of the Hodge star
operator on the volume-form one can express the variation of the
metric in terms of the variations of the co-frame.  Thus the
variations of the co-frame are related to the variations of the
brane world coordinates via the constraints. For this reason in
this new approach where we also consider the brane world metric as
a fundamental independent field one can not simply vary the action
in \eqref{e122.6} and equate the variation coefficients to zero to
find the field equations since they are not independent and they
are related by the constraint equations. One way of overcoming
this difficulty is to insert the constraint equation \eqref{e2} a
priori in the action of \eqref{e122.6} then vary it abolishing the
dependency of the variations. However still the resulting field
equations must be solved together with the constraints \eqref{e2}.
Now if we use the constraint equation of \eqref{e122.6} in the
action of \eqref{e122.6} the braneworld kinetic term can be added
to the cosmological constant term and we end up with the
constraint system
\begin{subequations}\label{e123}
\begin{align}
S_{Brane}&=\int_{W}\bigg(\;\frac{1}{16\kappa\pi}\;R_{AB}^{(W)}\wedge\ast
e^{AB}+\ast\Lambda_{eff}\bigg)+\int_{W}\mathcal{L}_{Matter},\notag\\
\notag\\
G_{AB}&=g_{ab}\partial_{A}
 \varphi^{a}\partial_{B}\varphi^{b}\tag{\ref{e123}},
 \end{align}
\end{subequations}
where $\Lambda_{eff}=\Lambda-T_{p}$ and the free fields are the
$\{G_{AB},\varphi^{a}\}$ and the matter fields. At this stage one
may simply vary the action and derive the field equations without
worrying about the constraints as we have mentioned above. Even,
since \eqref{e123} does not have any explicit $\{\varphi^{a}\}$
dependence if one suppresses and ignores the braneworld coordinate
degrees of freedom as it is usual in the relative literature one
may end up with an ordinary matter-gravity coupling problem on the
braneworld. In this methodology one may follow two tracks after
deriving the Einstein equations from \eqref{e123} which do not
have any explicit $\{\varphi^{a}\}$ dependence. Firstly if the
bulk metric is not imposed one may study the braneworld gravity
for various $G_{AB}$ and its spin connections and then by using
the constraint equations of \eqref{e123} one may lift the results
to the bulk. The second problem is the most general case which is
therefore more involved. In this case the bulk metric may be
assigned directly or it may emerge as a result of some dynamics in
the bulk. Then although not explicitly appearing in the Einstein
equations corresponding to \eqref{e123}, as hidden in the
constraint part of \eqref{e123} the braneworld coordinates
$\{\varphi^{a}\}$ link the bulk dynamics to the brane dynamics. In
this case since the bulk metric becomes a function of the
braneworld coordinates at the brane-bulk interface one looses the
freedom to start with an arbitrary braneworld metric $G_{AB}$ to
compute the relative spin connection. Therefore one must find
means to express the Einstein equations obtained from \eqref{e123}
in terms of the true-original independent fields of \eqref{e122}
which are the braneworld coordinates. In this manner the
braneworld dynamics can be directly related to the bulk one. The
direct way of expressing the braneworld gravity sector in terms of
the brane coordinates is to start with the constraint part of
\eqref{e123} and then to derive the associated Levi-Civita
connection on the bulk and to use the result in the Einstein
equation. As the reader may realize this is technically a very
non-linear method as we have mentioned before. The naturally
preferred method is to make use of the Gauss equation of Section
six, this is possible as the braneworld is intrinsically an
isometric immersion  in the bulk via the original definition of
its dynamics. In this way following the usual gravity formulation
of the braneworld which includes $R^{(W)}_{AB}$ one may directly
implement the $\{\varphi^{a}\}$ dependence via the Gauss equation
\eqref{e121} whose right hand side is a function of
$\{\varphi^{a}\}$. This will be the method we will use in this
section. Up to now the formulation we have discussed remains in
its most general form. Our major contribution within this general
formalism is to specify the bulk. In this section we will
specialize in the case where the bulk is a $G/K$ symmetric space.
In fact this is at the heart of the present work and although
Section six presents the general formalism of relating the bulk
and the braneworld, Sections four and five will provide the
special bulk metric and its bulk spin connection which are
required in the Gauss equation \eqref{e121}. Before deriving the
field equations in the above mentioned direction we should remark
one point. The Nambu-Goto action in \eqref{e122.6} can freely be
exchanged with the Polyakov one \eqref{e8} without needing a
further cosmological constant contribution. This is possible
because the Polyakov version of \eqref{e122.6} which is also a
constraint system will produce a non-vanishing energy-momentum
tensor as we have gravity and matter field couplings now. Thus the
constraint equations in \eqref{e122.6} are no more inconsistent
with the Einstein equation of this equivalent Polyakov system.
Thus they can be freely used to obtain the Nambu-Goto form. This
justifies the equivalence of the Nambu-Goto and the Polyakov
versions of \eqref{e122.6}\footnote{In this case we have
$T_{p}=T(p+1)$.}. We will not directly make use of the Polyakov
version but this equivalence will enable us to assume the special
form of the brane dynamics in the form of the symmetric space
sigma model at the Polyakov level which is discussed in detail in
Section three. Therefore we can state that our symmetry discussion
of Section three keeps its validity for \eqref{e122} and its
equivalents \eqref{e122.6} and \eqref{e123}. The bulk metric
$g_{ab}$ at the braneworld bulk intersection is again dictated by
the symmetries of the brane theory which we will assume to be
described by a symmetric space sigma model at the equivalent
Polyakov formalism. We assume the symmetric space sigma model
dynamics for the brane kinetic term so that in this way the brane
dynamics may exhibit a rich class of conserved local and global
charges which can reflect itself as a rich structure of
symmetries. The price one pays for this is the restrictions on the
isometric immersion of the world volume in the symmetric space
bulk. This is the major complication of this specific problem
which we have already faced within the computation of the bulk
curvature in Section five. As we have seen in Section four the
symmetry requirements of the braneworld kinetics fix the form of
the bulk metric at the braneworld intersection in terms of the
braneworld coordinates as \eqref{e43}. Thus in this special case
in \eqref{e122} and its equivalents \eqref{e122.6}, \eqref{e123}
the bulk metric becomes \eqref{e43}. With this observation finally
we can express the complete dynamics of the brane whose motion in
a symmetric space bulk is governed by a symmetric space sigma
model and which is coupled to the gravity and the matter fields as
\begin{subequations}\label{e123.5}
\begin{align}
S_{Brane}&=\int_{W}\bigg(\;\frac{1}{16\kappa\pi}\;R_{AB}^{(W)}\wedge\ast
e^{AB}+\ast\Lambda_{eff}\bigg)+\int_{W}\mathcal{L}_{Matter},\notag\\
\notag\\
G_{AB}&=g_{ab}\partial_{A}
 \varphi^{a}\partial_{B}\varphi^{b},\notag\\
 \notag\\
g&=-\frac{1}{8}\mathcal{A}_{ij}d\phi^{i}\otimes
d\phi^{j}-\frac{1}{8}\,\mathcal{B}_{i\alpha}
e^{\frac{1}{2}\alpha_{j}\phi^{j}}
\mathbf{\Omega}^{\alpha}_{\beta}d\phi^{i}\otimes
d\chi^{\beta}\notag\\
\notag\\
&\;\;\;\;-\frac{1}{8}\,\mathcal{B}_{i\alpha}
e^{\frac{1}{2}\alpha_{j}\phi^{j}}
\mathbf{\Omega}^{\alpha}_{\beta}d\chi^{\beta}\otimes d\phi^{i}
-\frac{1}{2}\,\mathcal{C}_{\alpha\beta}e^{\frac{1}{2}\alpha_{j}\phi^{j}}
e^{\frac{1}{2}\beta_{i}\phi^{i}}\mathbf{\Omega}^{\alpha}_{\gamma}
\mathbf{\Omega}^{\beta}_{\tau}d\chi^{\gamma}\otimes
d\chi^{\tau}.\tag{\ref{e123.5}}
 \end{align}
\end{subequations}
The later two are constraints on the independent fields
$\{\varphi^{a}(x^{A}),G_{AB}\}$. The method we will follow to
express the gravity sector of \eqref{e123.5} in terms of the true
independent fields which are the braneworld coordinates is in
three steps. First we will derive the field equations by
considering $G_{AB}$ as an independent field in addition to the
braneworld coordinates. This is possible due to the construction
of the action \eqref{e123.5} by introducing constraints as we have
discussed above. Secondly one computes the Levi-Civita connection
of the bulk metric given in \eqref{e123.5}. This is exactly what
we have done in Section five. We will simply adopt the results
from there. Thirdly in the Gauss equation we will use this bulk
spin connection arising from the braneworld symmetries at the
intersection of the bulk with the braneworld. When this form of
the Gauss equation is used in the isometrically immersed
braneworld curvature which enters into the braneworld Einstein
equation then it finally reformulates the dynamics in terms of the
braneworld coordinates. In all of the following formulation as the
reader may appreciate we will avoid using the explicit form of the
bulk spin curvature components derived in \eqref{e85},
\eqref{e93}, and \eqref{e94} to save space and to prevent our
results from looking excessively messy. For this reason in the
following formulas the bulk spin curvature components will be left
in their compact form. Therefore although the results are also
applicable for the general bulk metric the expressions when the
bulk curvature components are substituted from \eqref{e85},
\eqref{e93}, and \eqref{e94} explicitly computes the braneworld
gravity sector in terms of the braneworld coordinates for the
special case of the symmetric space bulk metric arising from the
symmetric space sigma model kinetics of the brane. Apart form the
complementing derivations of sections four and five these
equations\footnote{Though left in an implicit form.} defining the
braneworld gravity in symmetric space bulk is the main objective
of this work. Now if we vary the action in \eqref{e123.5} we get
\begin{subequations}\label{e124}
\begin{align}
 \delta S_{Brane}&=\int_{W}\bigg(\;\frac{1}{16\kappa\pi}\;\delta R_{AB}^{(W)}\wedge\ast
e^{AB}+\frac{1}{16\kappa\pi}\;R_{AB}^{(W)}\wedge\delta\ast
e^{AB}+\Lambda_{eff}\delta\ast 1\bigg)\notag\\
\notag\\
&\:\:\:\:+\int_{W}\delta\mathcal{L}_{Matter}\notag\\
\notag\\
&=\int_{W}\delta e^{C}\wedge\bigg(\frac{1}{16\kappa\pi}\;\ast
e_{CAB}\wedge R^{(W)\:AB}+\Lambda_{eff}\ast e_{C}+\ast
t_{C}\bigg)\notag\\
\notag\\
&\:\:\:\:+\int_{W}F(\delta X_{other}),\tag{\ref{e124}}
 \end{align}
\end{subequations}
where the first term on the RHS vanishes because it can be written
in terms of $\delta\omega^{A}_{\:B}$ or $\delta\Gamma^{A}_{BC}$
which are functions of the variations of the torsion components
which vanish as we assume that the connection is the Levi-Civita
connection on the braneworld \cite{wheeler}. Alternatively
\cite{thring} if one assumes that $\{e^{A}\}$ is an orthogonal
frame that term can be shown to generate a surface term which
vanishes again due to the assumption of the vanishing variations
at the boundaries. In \eqref{e124} $\{t_{C}\}$ are the braneworld
energy-momentum one-forms which result from the metric related
variation terms in $\delta\mathcal{L}_{Matter}$ and $\{\delta
X_{other}\}$ are the non-metric related variations of the other
fields in the theory. Now if we equate \eqref{e124} to zero we get
the Einstein equation for the braneworld as
\begin{equation}\label{e125}
-\frac{1}{16\kappa\pi}\;\ast e_{CAB}\wedge
R^{(W)\:AB}-\Lambda_{eff}\ast e_{C}=\ast t_{C}.
\end{equation}
In terms of the braneworld Riemann tensor components this equation
can be written as \cite{thring}
\begin{equation}\label{e126}
\bigg(-\frac{1}{16\kappa\pi}\;
R^{(W)A\:\:\:\:\:\:B}_{\:\:\:\:\:\:\:\:\:\:\:\:BA}G_{CD}+\frac{1}{8\kappa\pi}\;
R^{(W)A}_{\:\:\:\:\:\:\:\:\:\:\:\:CAD}-\Lambda_{eff}G_{CD}\bigg)\ast
e^{D}=T_{CD}\ast e^{D},
\end{equation}
where $T_{CD}e^{D}=t_{C}$ and $T_{CD}$ is the energy-momentum
tensor of the matter fields on the braneworld. Thus the Einstein
equation in its familiar form reads
\begin{equation}\label{e127}
-\frac{1}{16\kappa\pi}\;
R^{(W)A\:\:\:\:\:\:B}_{\:\:\:\:\:\:\:\:\:\:\:\:BA}G_{CD}+\frac{1}{8\kappa\pi}\;
R^{(W)A}_{\:\:\:\:\:\:\:\:\:\:\:\:CAD}-\Lambda_{eff}G_{CD}=T_{CD}.
\end{equation}
This equation together with the matter field equations must be
simultaneously solved with the constraints appearing in
\eqref{e123.5}. For this reason the constraint equations must be
implemented into the Einstein equation \eqref{e127}. Although a
direct substitution can be considered as a first guess it will
algebraically be more involved and geometrically less apparent.
Alternatively as we have discussed we will make use of the Gauss
equation of the last section which reveals the immersion geometry
characteristics by expressing the equations explicitly in terms of
the extrinsic curvature. In this manner, one can also monitor the
geometry of the braneworld inside the bulk, besides one can easily
tune further restrictions on this geometry within such a form of
Einstein equation. Especially this explicit form of the gravity
dynamics can be directly coupled to the bulk dynamics via the
braneworld coordinates. Now let us consider the straightforward
substitution of the constraint equation  in \eqref{e127}. Firstly
\cite{wheeler,schutz}
\begin{equation}\label{e128}
R^{(W)A\:\:\:\:\:\:}_{\:\:\:\:\:\:\:\:\:\:\:\:BCD}=\tilde{\Gamma}^{E}_{DB}\tilde{\Gamma}^{A}_{CE}-
\tilde{\Gamma}^{E}_{CB}\tilde{\Gamma}^{A}_{DE}+\partial_{C}\tilde{\Gamma}^{A}_{DB}-\partial_{D}\tilde{\Gamma}^{A}_{CB},
\end{equation}
where we have used the fact that $\{r_{A}\}=\{\partial/\partial
x^{A}\}$. The Christoffel coefficients can be calculated from the
induced braneworld metric components as
\begin{subequations}\label{e129}
\begin{align}
\tilde{\Gamma}_{AB}^{C}&=\frac{1}{2}\:g_{ab}\partial^{C}\varphi^{a}\partial^{D}\varphi^{b}
\big[\partial_{A}(g_{ab}\partial_{B}\varphi^{a}\partial_{D}\varphi^{b})
+\partial_{B}(g_{ab}\partial_{D}\varphi^{a}\partial_{A}\varphi^{b})\notag\\
\notag\\
&\:\:\:\:-\partial_{D}(
g_{ab}\partial_{A}\varphi^{a}\partial_{B}\varphi^{b})\big]\tag{\ref{e129}},
\end{align}
\end{subequations}
where $g_{ab}$ is the bulk metric in \eqref{e123.5}. Now if one
substitutes \eqref{e128} into \eqref{e127} one gets
\begin{subequations}\label{e130}
\begin{align}
 &-\frac{1}{16\kappa\pi}\;(g_{ab}\partial_{C}\varphi^{a}\partial_{D}\varphi^{b})
 \bigg[\tilde{\Gamma}^{EB}_{\:\:\:\:\:\:B}\tilde{\Gamma}^{A}_{AE}-
\tilde{\Gamma}^{E}_{AB}\tilde{\Gamma}^{AB}_{\:\:\:\:\:\:E}
+\partial_{A}\tilde{\Gamma}^{AB}_{\:\:\:\:\:\:B}-\partial^{B}\tilde{\Gamma}^{A}_{AB}\bigg]\notag\\
\notag\\
&+\frac{1}{8\kappa\pi}\;\bigg[\tilde{\Gamma}^{E}_{DC}\tilde{\Gamma}^{A}_{AE}-
\tilde{\Gamma}^{E}_{AC}\tilde{\Gamma}^{A}_{DE}+\partial_{A}\tilde{\Gamma}^{A}_{DC}
-\partial_{D}\tilde{\Gamma}^{A}_{AC}\bigg]-\Lambda_{eff}(g_{ab}\partial_{C}
\varphi^{a}\partial_{D}\varphi^{b})\notag\\
\notag\\
&=T_{CD}. \tag{\ref{e130}}
 \end{align}
\end{subequations}
It is obvious that though direct, this equation is insensitive to
the geometrical characteristics of the immersion. For this reason
beside \eqref{e130} we will also present another formulation
combining the constraint equation and the Einstein equation by
means of the Gauss equation. By this way, one can explicitly
monitor the bulk and the braneworld curvature relationship. Such a
form may find more use when one also adds bulk gravity to the
theory. The Bulk curvature on the braneworld is calculated in
Section five with respect to a general bulk moving co-frame
\eqref{e54} which is obtained from the bulk coordinate frame by
dressings. However the Gauss equation \eqref{e121} which is
certainly dependent on the bulk frame chosen (and thus a component
equation) is constructed with respect to the basis \eqref{e101}
which is rather special and which enables one to derive
\eqref{e121} as it is based on the geometrical decomposition
\eqref{e98}. Thus to be able to use the Riemann tensor components
belonging to the bulk curvature components \eqref{e85},
\eqref{e93}, and \eqref{e94} in the Gauss equation \eqref{e121} we
must do two steps of bulk co-frame transformations from
\eqref{e54} to \eqref{e101} and the associated bulk curvature
transformations. Namely starting from the ones in \eqref{e85},
\eqref{e93}, and \eqref{e94} we must reach the ones used in
\eqref{e121} which correspond to a special moving co-frame. In
\eqref{e85}, \eqref{e93}, and \eqref{e94}, we have computed the
bulk curvature two-forms
 with respect to the
moving co-frame $\{e^{a}\}$ which is related to the coordinate
co-frame $\{d\varphi^{a}\}$ as
\begin{equation}\label{e131}
e^{a}=L^{a}_{\:\:\:b}d\varphi^{b},
\end{equation}
where
\begin{equation}\label{e132}
L=\overset{\scriptstyle i\scriptstyle
=\scriptstyle1,\scriptstyle2,\scriptstyle\cdots\scriptstyle\cdots\scriptstyle\cdot\scriptstyle
r\quad\quad\quad\scriptstyle \alpha\scriptstyle
=\scriptstyle1,\scriptstyle2,\scriptstyle\cdots\scriptstyle\cdots\scriptstyle\cdot\scriptstyle
n}
{\left(\begin{array}{c|c} \begin{pmatrix} &&&&\\
&&\text{\Large{1}}&&\\&&&&
\end{pmatrix}&
\begin{pmatrix}&&&&\\
&&\text{\Large{0}}&&\\&&&&\end{pmatrix}    \\
\hline
\begin{pmatrix}&&&&\\&&\text{\Large{0}}&&\\&&&&\end{pmatrix}&
\begin{pmatrix}&&&&\\&&\text{\Large{$\tilde{\Omega}$}}&&\\&&&&\end{pmatrix}
\end{array}\right)}\begin{array}{c}\overset{\overset{\underset{\scriptstyle\shortparallel}{\scriptstyle
i}}{ \scriptstyle 1}}{\underset{\scriptstyle
r}{\overset{\scriptstyle\cdot}{\overset{\scriptstyle\cdot}{\overset{\scriptstyle\cdot}{\scriptstyle\cdot}
}}}}\\\overset{\overset{\underset{\scriptstyle\shortparallel}{\scriptstyle
\alpha}}{ \scriptstyle 1}}{\underset{\scriptstyle
n}{\overset{\scriptstyle\cdot}{\overset{\scriptstyle\cdot}{\overset{\scriptstyle\cdot}{\scriptstyle\cdot}
}}}}\\\\\end{array}.
\end{equation}
Here
\begin{equation}\label{e133}
\tilde{\Omega}^{\alpha}_{\:\:\:\beta}=e^{\frac{1}{2}\alpha_{j}\phi^{j}}
\Omega^{\alpha}_{\:\:\:\beta}.
\end{equation}
The dual frame transforms as
\begin{equation}\label{e134}
b_{c}=\frac{\partial}{\partial\varphi^{a}}(L^{-1})^{a}_{\:\:\:c}.
\end{equation}
By using the transformation \eqref{e134} one can compute the
curvature two-forms (the ones with the check below) with respect
to the bulk coordinate frame $\{\partial/\partial\varphi^{a}\}$ as
\cite{thring}\footnote{One should be careful when writing
\eqref{e135} as one needs to perform index raising in \eqref{e85},
\eqref{e93}, and \eqref{e94} to construct \eqref{e135} and here or
anywhere else in this manuscript one has to use the appropriate
bulk or braneworld metric components which correspond to the frame
being worked in.}
\begin{equation}\label{e135}
\check{R}^{a}_{\:\:\:b}=(L^{-1})^{a}_{\:\:\:d}R^{d}_{\:\:\:e}L^{e}_{\:\:\:b},
\end{equation}
from which one can read the Riemann tensor components as
\begin{equation}\label{e136}
\check{R}_{ab}=\frac{1}{2}\check{R}_{abcd}d\varphi^{c}\wedge
d\varphi^{d}.
\end{equation}
In order to make use of the Riemann tensor components emerging
form the computed curvature elements \eqref{e85}, \eqref{e93}, and
\eqref{e94} in the Gauss equation \eqref{e121} we have to do one
more transformation on them to relate them to the ones appearing
in \eqref{e121}. This time we have to transform the Riemann tensor
components from the coordinate basis
$\{\partial/\partial\varphi^{a}\}$ to the one $\{e^{\prime}_{a}\}$
given in \eqref{e101}. These are related via \eqref{e101.7} as
\begin{equation}\label{e137}
e^{\prime}_{a}=\frac{\partial}{\partial\varphi^{b}}S^{b}_{\:\:\:a},
\end{equation}
where we know that $S^{b}_{\:\:\:A}=\partial_{A}\varphi^{b}$ thus
\begin{equation}\label{e138}
r_{A}=\frac{\partial}{\partial\varphi^{b}}\partial_{A}\varphi^{b}.
\end{equation}
By using this transformation now we can express the Riemann tensor
components $R_{ABCD}$ which are with respect to the frame
$\{e^{\prime}_{a}\}$ for the indices
$a,b,\cdots=1,\cdots,p+1$\footnote{That is to say the components
for the subset of indices $a,b,\cdots\equiv A,B,\cdots$.} in terms
of the ones with respect to the frame
$\{\partial/\partial\varphi^{a}\}$. The former are the ones on
which the Gauss equation is based on and the later are the
explicitly computed ones and they can be read from \eqref{e85},
\eqref{e93}, and \eqref{e94} via \eqref{e135}, \eqref{e136}. Thus
from \eqref{e137} and \eqref{e138} we have
\begin{equation}\label{e139}
R_{ABCD}=\check{R}_{abcd}\partial_{A}\varphi^{a}\partial_{B}\varphi^{b}
\partial_{C}\varphi^{c}\partial_{D}\varphi^{d}.
\end{equation}
Finally we can write the Gauss equation
\eqref{e121}\footnote{Though notationally implicit as we have
discussed before the reader should consider the following
equations together with the equations \eqref{e85}, \eqref{e93},
\eqref{e94}, \eqref{e135}, and \eqref{e136} which lead to the
explicit expressions of the Einstein equation in terms of the
braneworld coordinates.} explicitly in terms of the braneworld
coordinates as
\begin{equation}\label{e140}
R^{W}_{ABCD}=\check{R}_{abcd}\partial_{A}\varphi^{a}\partial_{B}\varphi^{b}
\partial_{C}\varphi^{c}\partial_{D}\varphi^{d}+\mathcal{F}_{AC}^{\:\:\:\:\:\:m}\mathcal{F}_{BDm}
-\mathcal{F}_{BC}^{\:\:\:\:\:\:m}\mathcal{F}_{ADm}.
\end{equation}
Substitution of this relation into the braneworld Einstein
equation \eqref{e127} finalizes the process of implementing the
induced braneworld metric constraint and deriving the gravity
dynamics explicitly in terms of the braneworld coordinates which
enter into the following equations through the explicitly computed
bulk curvatures of Section five which we leave in compact form.
Thus finally we have
\begin{subequations}\label{e141}
\begin{align}
 &-\frac{1}{16\kappa\pi}\;(g_{ef}\partial_{C}\varphi^{e}\partial_{D}\varphi^{f})
 \bigg[\check{R}_{abcd}\partial^{A}\varphi^{a}\partial_{B}\varphi^{b}
\partial_{A}\varphi^{c}\partial^{B}\varphi^{d}+\mathcal{F}_{\:\:\:A}^{A\:\:m}
\mathcal{F}^{\:\:\:B}_{B\:\:\:m}\notag\\
\notag\\
&-\mathcal{F}_{BA}^{\:\:\:\:\:\:m}\mathcal{F}^{AB}_{\:\:\:\:\:\:\:m}\bigg]+\frac{1}{8\kappa\pi}\;\bigg[\check{R}_{abcd}\partial^{A}\varphi^{a}\partial_{C}\varphi^{b}
\partial_{A}\varphi^{c}\partial_{D}\varphi^{d}+\mathcal{F}_{\:\:\:A}^{A\:\:m}\mathcal{F}_{CDm}\notag\\
\notag\\
&-\mathcal{F}_{CA}^{\:\:\:\:\:m}\mathcal{F}^{A}_{\:\:\:Dm}\bigg]-\Lambda_{eff}(g_{ab}\partial_{C}
\varphi^{a}\partial_{D}\varphi^{b})=T_{CD}. \tag{\ref{e141}}
 \end{align}
\end{subequations}
We remark that here the indices $A,B,\cdots$ are raised and
lowered by the induced braneworld metric \eqref{e2} whose
components are computed with respect to the braneworld frame
$\{r_{A}\}=\{\partial/\partial x^{A}\}$. In addition to being
directly in terms of the true independent fields of the theory
since expressed in terms of the second fundamental form or the
extrinsic curvature components as tuning parameters of geometry it
must be apparent to the reader that geometrically \eqref{e141} is
more eligible than \eqref{e130}. One straightforward application
of managing the geometry of the braneworld immersion through
\eqref{e141} can be said to occur when the braneworld immersion is
desired to be a totally geodesic one. In this case one demands
that any geodesic in the braneworld is also a geodesic in the bulk
with respect to the bulk metric connection. This requirement can
simply be considered to emerge from a physical need of the
completeness of the gravity theory; the geodesic motion must be
valid both in the bulk and the braneworld. In this case to
generate such solutions, in \eqref{e141} one may simply equate the
$\mathcal{F}$-terms to zero as for totally geodesic immersions the
second fundamental form is identically zero sufficiently and
necessarily \cite{aminov,docarmo,sternberg,oneil}. However to
generate the entire solution space of such totally geodesic
immersions one must introduce the vanishing of the second
fundamental form as a constraint equation at the level of total
action.
\section{Conclusion}
After discussing the low energy free Dp-brane dynamics in
symmetric space $G/K$ bulk and comparing the Nambu-Goto and the
Polyakov actions in Section two we introduced the symmetric space
sigma model action with the global $G$ and the local $K$ symmetry
in the solvable lie algebra gauge in Section three. By inspecting
this action in Section four, we have read the bulk metric which is
required at the symmetric space bulk-braneworld intersection for
the symmetries to occur. We have also discussed that this bulk
metric which we have explicitly constructed on a local coordinate
chart must be $G$-invariant. In the following section, we have
calculated the elements of the Levi-Civita connection
corresponding to this metric on the bulk. Later, following the
construction of the Gauss equation of the braneworld immersion of
Section six, in Section seven we have discussed the implementation
of the first fundamental form into the Einstein equation when the
induced braneworld metric is coupled to gravity as the brane moves
in the symmetric space bulk. Although the finalizing formulae of
Section seven are left in their most compact form as we have
discussed several times when incorporated with the explicit
curvature computations of Section five they lead us to the
expressions of the brane dynamics purely in terms of the
braneworld coordinates. This perspective which is aimed from the
beginning sits at the core of the entire formulation.

In this work, we have presented the basics of the braneworld
gravity when the brane motion takes place in a symmetric space.
Due to the symmetry properties of the bulk which can be
systematically constructed from Lie groups the problem of brane
motion in symmetric space has restrictive and governing
characteristics. In this manner, specifying the nature of the bulk
has enabled us to refine the general braneworld gravity analysis
to the particular case studied in this work. We have focussed on
the symmetry characteristics of the problem which require a
special form of a bulk metric in terms of the solvable lie algebra
gauge parameters of the sigma model. In a series of sections our
primary point of view was to compute the basic steps of the
standard braneworld gravity \cite{grav1} which lead to the
substitution of the Gauss equation into the Einstein equation for
our specially required bulk metric which is put on the scene by
the symmetries of the brane action. On the contrary, we have not
examined and discussed the details of the physical aspects of the
problem \cite{grav1,grav2,wald}. We can say that, the main concern
of this work is to study the technically somewhat involved
immersion structure of the braneworld gravity when the background
is a symmetric space. In this direction we have studied and shed
light on the Gauss and Einstein equation correspondence of
braneworld gravity and cosmology for symmetric space backgrounds.
Therefore the results of this work are computational rather than
being physical. However, starting from the results achieved here,
by introducing also the bulk gravity one can study further the
junction conditions, the initial value problem, the
brane-observers etc. The details in these directions need the
further manipulation of the relations of the bulk and the
braneworld curvatures in the Einstein equations which may be
obtained from the structural equations of the braneworld immersion
likewise the Codazzi equation. Furthermore, the specification of
the bulk and the braneworld matter fields may enable one to study
the physical solutions and the braneworld geometries including the
FRW braneworld cosmologies in symmetric space bulk. By this way
one can link the connection to the RS braneworld models
\cite{ran1,ran2,bran1,bran2,bran3,bran4,bran5}. Although we have
presented the basic equations of the braneworld immersion gravity,
our formulation, in its general form is for an arbitrary
dimensional braneworld in a generic symmetric space background.
Therefore on its own right the identification of the $G-$invariant
bulk metric and the calculation of its Levi-Civita connection
contributes new computational results to the geometry of symmetric
spaces. On the other hand, the reader should appreciate the
relevance of the special kind of bulk chosen here to the
M-theoretical holography principle
\cite{hol1,hol2,hol3,hol4,hol4.5,hol4.6}. Studying braneworld
gravity in symmetric space bulk can form a new computational
region in which one can search for new realizations of the
M-theoretical holography principle
\cite{hol1,hol2,hol3,hol4,hol4.5,hol4.6} within the context of
classical and braneworld cosmologies
\cite{hol5,hol6,hol7,hol8,hol9,hol9.5,hol10,hol11}.


\begin{thebibliography}{99}
\bibitem{ran1}
L. Randall and R. Sundrum, ``\textit{An alternative to
compactification}", Phys. Rev. Lett. \textbf{83} (1999) 4690,
hep-th/9906064.
\bibitem{ran2}
L. Randall and R. Sundrum, ``\textit{A large mass hierarchy from a small extra dimension}",
Phys. Rev. Lett.  \textbf{83} (1999) 3370, hep-ph/9905221.
\bibitem{bran1}
D. Langlois, ``\textit{Brane cosmology: An introduction}",
Prog. Theor. Phys. Suppl. \textbf{148} (2003) 181, hep-th/0209261.
\bibitem{bran2}
J. E. Lidsey, ``\textit{Inflation and braneworlds}",
Lect. Notes Phys. \textbf{646} (2004) 357, astro-ph/0305528.
\bibitem{bran3}
A. O. Barvinsky, ``\textit{Braneworld effective action and origin of inflation}",
Phys. Rev. \textbf{D65} (2002) 062003, hep-th/0107244.
\bibitem{bran4}
G. Huey and J. E. Lidsey, ``\textit{Inflation, braneworlds and quintessence}",
Phys. Lett. \textbf{B514} (2001) 217, astro-ph/0104006.
\bibitem{bran5}
N. T. Jones, H. Stoica and S. H. H. Tye, ``\textit{Brane
interaction as the origin of inflation}", JHEP \textbf{0207}
(2002) 051, hep-th/0203163.
\bibitem{grav1}
R. Maartens, ``\textit{Brane-world gravity}", Living Rev. Rel.
\textbf{7} (2004) 7, gr-qc/0312059.
\bibitem{hel}
S. Helgason, ``\textbf{Differential Geometry, Lie Groups and
Symmetric Spaces}", (Graduate Studies in Mathematics, 34, American
Mathematical Society Providence R. I., 2001).
\bibitem{pbrane1}
C. V. Johnson, ``\textit{D-brane primer}", hep-th/0007170.
\bibitem{pbrane2}
J. A. Garcia, R. Linares and J.~D. Vergara, ``\textit{Weyl
invariant p-brane and Dp-brane actions}", Phys. Lett. {\textbf
B503} (2001) 154, hep-th/0011085.
\bibitem{pbrane3}
G. W. Gibbons, ``\textit{Born-Infeld particles and Dirichlet
p-branes}", Nucl. Phys. {\textbf B514} (1998) 603, hep-th/9709027.
\bibitem{pbrane4}
M. J. Duff, ``\textit{Supermembranes}", hep-th/9611203.
\bibitem{pbrane5}
M. Cederwall, ``\textit{Aspects of D-brane actions}", Nucl. Phys.
Proc. Suppl. {\textbf 56B} (1997) 61, hep-th/9612153.
\bibitem{pbrane6}
J. Polchinski, ``\textit{Lectures on D-branes}", hep-th/9611050.
\bibitem{pbrane7}
M. Abou Zeid and C. M. Hull, ``\textit{Intrinsic geometry of
D-branes}", Phys. Lett. {\textbf B404} (1997) 264, hep-th/9704021.
\bibitem{pbrane8}
M. Abou Zeid and C. M. Hull, ``\textit{Geometric actions for
D-branes and M-branes}", Phys. Lett. {\textbf B428} (1998) 277,
hep-th/9802179.
\bibitem{howe}
P. S. Howe and R. W. Tucker, ``\textit{A locally supersymmetric
and reparametrization invariant action for a spinning membrane}",
J. Phys. {\textbf A10} (1977) L155.
\bibitem{polyakov}
A. M. Polyakov, ``\textit{Quantum geometry of bosonic strings}",
Phys. Lett. {\textbf B103} (1981) 207.
\bibitem{sm1}
H. Eichenherr and M. Forger, ``\textit{On the dual symmetry of the
nonlinear sigma models}", Nucl. Phys. \textbf{B155} (1979) 381.
\bibitem{sm2}
H. Eichenherr and M. Forger, ``\textit{More about nonlinear sigma
models on symmetric spaces}", Nucl. Phys. \textbf{B164} (1980)
528.
\bibitem{sm3}
H. Eichenherr and M. Forger, ``\textit{Higher local conservation
laws for nonlinear sigma models on symmetric spaces}", Commun.
Math. Phys. \textbf{82} (1981) 227.
\bibitem{Eichen}
H. Eichenherr, ``\textit{Geometric analysis of integrable
nonlinear sigma models}", Lect. Notes Phys. \textbf{151} (1982)
189.
\bibitem{zak}
V. E. Zakharov and A. V. Mikhailov, ``\textit{Relativistically
invariant two-dimensional models of field theory which are
integrable by means of the inverse scattering problem method}",
Sov. Phys. JETP \textbf{47} (1978) 1017; Zh. Eksp. Teor. Fiz.
\textbf{74} (1978) 1953.
\bibitem{uhlen}
K. Uhlenbeck, ``\textit{Harmonic maps into Lie groups: Classical
solutions of the chiral model}", J. Diff. Geo. \textbf{30} (1989)
1.
\bibitem{manas}
A. P. Fordy (Ed.) and J. C. Wood (Ed.), ``\textbf{Harmonic Maps
and Integrable Systems}", (Aspects of Mathematics, Vol. E23,
Vieweg, Braunschweig/Wiesbaden, 1994).
\bibitem{julia1}
E. Cremmer, B. Julia,
H. L\"{u} and C. N. Pope, ``\textit{Dualisation of dualities.
I.}", Nucl. Phys. \textbf{B523} (1998) 73, hep-th/9710119.
\bibitem{julia2}
 E. Cremmer, B. Julia, H. L\"{u} and C. N. Pope, ``\textit{Dualisation of dualities II : Twisted self-duality of doubled fields and
 superdualities}",
Nucl. Phys. \textbf{B535} (1998) 242, hep-th/9806106.
\bibitem{gauged}
R. Percacci and E. Sezgin, ``\textit{Properties of gauged sigma
models}", (Published in College Station 1998, Relativity, particle
physics and cosmology 255-278), hep-th/9810183.
\bibitem{nej1}
N. T. Y$\i$lmaz, ``\textit{Dualisation of the general scalar coset
in supergravity theories}", Nucl. Phys. \textbf{B664} (2003) 357,
hep-th/0301236.
\bibitem{nej2}
N. T. Y$\i$lmaz, ``\textit{The non-split scalar coset in
supergravity theories}", Nucl. Phys. \textbf{B675} (2003) 122,
hep-th/0407006.
\bibitem{nej3}
T. Dereli and N. T. Y$\i$lmaz, ``\textit{Dualisation of the
symmetric space sigma model with couplings}", Nucl. Phys.
\textbf{B705} (2005) 60, hep-th/0507007.
\bibitem{sssm1}
N. T. Y$\i$lmaz, ``\textit{Symmetric space sigma-model dynamics:
Internal metric formalism}", Phys. Lett. {\textbf B642} (2006)
270, hep-th/0611010.
\bibitem{born}
M. Born and L. Infeld, ``\textit{Foundations of the new field
theory}", Proc. Roy. Soc. Lond. {\textbf A144} (1934) 425.
\bibitem{dirac}
P. A. M. Dirac, ``\textit{An extensible model of the electron}",
Proc. Roy. Soc. Lond. {\textbf A268} (1962) 57.
\bibitem{darling}
R. W. R. Darling, ``\textbf{Differential Forms and Connections}",
(Cambridge University Press, 1995).
\bibitem{aminov}
Y. Aminov, ``\textbf{The Geometry of Submanifolds}", (Gordon and
Breach Science Pub. Amsterdam, 2001).
\bibitem{docarmo}
M. P. Do Carmo, ``\textbf{Riemannian Geometry}", (Birkhauser,
Boston, 1992).
\bibitem{sternberg}
S. Sternberg, ``\textbf{Semi-Riemann Geometry and General
Relativity}", (Lecture Notes, Unpublished, 2003).
\bibitem{oneil}
B. O'Neill, ``\textbf{Semi-Riemannian Geometry with Applications
to Relativity}", (Academic Press, San Diego, 1983).
\bibitem{sch1}
J. H. Schwarz, ``\textit{Classical symmetries of some
two-dimensional models}",
  Nucl. Phys. \textbf{B447} (1995) 137, hep-th/9503078.
\bibitem{fre}
L. Andrianopoli, R. D'Auria, S. Ferrara, P. Fr\'{e} and M.
Trigiante, ``\textit{RR scalars, U-duality and solvable Lie
algebras}", Nucl. Phys. \textbf{B496} (1997) 617, hep-th/9611014.
\bibitem{symmspace}
N. T. Y$\i$lmaz, ``\textit{On the symmetric space sigma-model
kinematics}", Int. J. Mod. Phys. \textbf{A22} (2007) 2683,
arXiv:0707.2150 [hep-th].
\bibitem{west}
P. C. West, ``\textit{Supergravity, brane dynamics and string
duality}", hep-th/9811101.
\bibitem{tanii}
 Y. Tanii, ``\textit{Introduction to supergravities in diverse dimensions}", hep-th/9802138.
\bibitem{cornwell}
J. F. Cornwell, ``\textbf{Group Theory in Physics: An
Introduction}", (Academic Press, San Diego, 1997).
\bibitem{kunze}
 K. M. Hoffman, R. A. Kunze, ``\textbf{Linear Algebra}",
 (2nd Edition, Prentice Hall, New Jersey, 1971).
\bibitem{isham}
 C. J. Isham, ``\textbf{Modern Differential Geometry for Physicists}",
(World Scientific Pub. Co., Singapore, 1989).
\bibitem{tucker}
I. M. Benn, R. W. Tucker, ``\textbf{An Introduction to Spinors and
Geometry with Applications in Physics}", (Adam Hilger, Bristol,
1987).
\bibitem{thring}
W. Thirring, ``\textbf{A Course in Mathematical Physics I and II:
Classical Dynamical Systems and Classical Field Theory}",
(Springer-Verlag, New York, 1992).
\bibitem{nakahara}
M. Nakahara, ``\textbf{Geometry, Topology and Physics}", (Adam
Hilger, Bristol, 1991).
\bibitem{wheeler}
C. W. Misner, K. S. Thorne and J. A. Wheeler,
``\textbf{Gravitation}", (W. H. Freeman, San Francisco 1973).
\bibitem{chern}
S. S. Chern, ``\textbf{Minimal Submanifolds in a Riemanman
Manifold}", (Technical Report 19, University of Kansas, Kansas,
1968).
\bibitem{lawson}
B. Lawson, ``\textbf{Lectures on Minimal Submanifolds}", (Vol. 1,
Publish or Perish, Berkeley, 1980).
\bibitem{schutz}
B. F. Schutz, ``\textbf{A First Course in General Relativity}",
(Cambridge University Press, 1985).
\bibitem{grav2}
T. Shiromizu, K. Maeda and M. Sasaki, ``\textit{The Einstein
equations on the 3-brane world}", Phys. Rev. \textbf{D62} (2000)
024012, gr-qc/9910076.
\bibitem{wald}
R. M. Wald, ``\textbf{General Relativity}", (University Of Chicago
Press, 1984).
\bibitem{hol1}
L. Susskind, ``\textit{The world as a hologram}'',
  J. Math. Phys. \textbf{36} (1995) 6377, hep-th/9409089.
\bibitem{hol2}
  C. R. Stephens, G. 't Hooft and B. F. Whiting,
  ``\textit{Black hole evaporation without information loss}",
  Class. Quant. Grav.  \textbf{11} (1994) 621, gr-qc/9310006.
\bibitem{hol3}
  E. Witten, ``\textit{Anti-de Sitter space and holography}",
  Adv. Theor. Math. Phys. \textbf{2} (1998) 253, hep-th/9802150.
\bibitem{hol4}
 L. Susskind and E. Witten,
  ``\textit{The holographic bound in anti-de Sitter space}", hep-th/9805114.
  \bibitem{hol4.5}
  J. M. Maldacena, ``\textit{The large N limit of superconformal field theories and
supergravity}",
  Adv. Theor. Math. Phys. \textbf{2} (1998) 231;
  Int. J. Theor. Phys. \textbf{38} (1999) 1113, hep-th/9711200.
  \bibitem{hol4.6}
  S. S. Gubser, I. R. Klebanov and A. M. Polyakov, ``\textit{Gauge theory correlators from
  non-critical string theory}",
  Phys. Lett. \textbf{B428} (1998) 105, hep-th/9802109.
\bibitem{hol5}
  N. Arkani-Hamed, M. Porrati and L. Randall, ``\textit{Holography and phenomenology}",
  JHEP \textbf{0108} (2001) 017, hep-th/0012148.
\bibitem{hol6}
  V. Balasubramanian, J. de Boer and D. Minic,
  ``\textit{Exploring de Sitter space and holography}",
  Class. Quant. Grav.  \textbf{19} (2002) 5655; Annals Phys. \textbf{303} (2003)
  59, hep-th/0207245.
  \bibitem{hol7}
  S. S. Gubser, ``\textit{AdS/CFT and gravity}",
  Phys. Rev. \textbf{D63} (2001) 084017, hep-th/9912001.
\bibitem{hol8}
  I. Savonije and E. P. Verlinde, ``\textit{CFT and entropy on the brane}",
  Phys. Lett. \textbf{B507} (2001) 305, hep-th/0102042.
  \bibitem{hol9}
  A. Padilla, ``\textit{Brane world cosmology and holography}", hep-th/0210217.
  \bibitem{hol9.5}
  W. Fischler and L. Susskind, ``\textit{Holography and cosmology}", hep-th/9806039.
\bibitem{hol10}
  D. Bak and S. J. Rey, ``\textit{Cosmic holography}",
  Class. Quant. Grav. \textbf{17} (2000) L83, hep-th/9902173.
  \bibitem{hol11}
  R. Bousso, ``\textit{Holography in general space-times}",
  JHEP \textbf{9906} (1999) 028, hep-th/9906022.
\end{thebibliography}
\end{document}